\documentclass[english,a4paper,11pt]{article}
\usepackage[english]{babel}
\usepackage[utf8]{inputenc}
\usepackage{cite}
\usepackage[affil-it]{authblk}
\usepackage{physics,stackrel}
\usepackage[dvipsnames]{xcolor}
\usepackage{slashed}
\usepackage{mathtools}
\usepackage{color}
\usepackage{siunitx}
\usepackage{simpler-wick,slashed}
\usepackage{graphicx}
\usepackage{jheppub}
\usepackage[nameinlink]{cleveref}

\graphicspath{{Figures/}}

\newcommand{\MSbar}{\ensuremath{\overline{\text{MS}}}}
\newcommand{\mbmb}{\ensuremath{\overline{m}_b \left(\overline{m}_b\right)}}
\newcommand{\mcmc}{\ensuremath{\overline{m}_c \left(\overline{m}_c\right)}}

\hypersetup{%
  colorlinks=true, linktocpage=true, pdfstartpage=1, pdfstartview=FitV,
  breaklinks=true, pageanchor=true,
  pdfpagemode=UseNone,
  plainpages=false, bookmarksnumbered, bookmarksopen=true, bookmarksopenlevel=1,
  hypertexnames=true, pdfhighlight=/O,
  urlcolor=RoyalBlue, linkcolor=ForestGreen, citecolor=RoyalBlue,
  pdftitle={HQET sum rules for matrix elements of dimension-six four-quark operators  for meson lifetimes  within and beyond the Standard Model},
  pdfauthor={Matthew Black, Martin Lang, Alexander Lenz, Zachary W\"uthrich},
  pdfsubject={},
  pdfkeywords={},
  pdfcreator={pdfLaTeX},
  pdfproducer={LaTeX with hyperref}
}

\title{HQET sum rules for matrix elements of dimension-six four-quark operators  for meson lifetimes  within and beyond the Standard Model}

\preprint{P3H-24-098, SI-HEP-2024-29}

\author[a,b]{Matthew Black,}
\author[a]{Martin Lang,}
\author[a]{Alexander Lenz,}
\author[a]{Zachary Wüthrich}
\affiliation[a]{Theoretische Teilchenphysik, Center for Particle Physics Siegen, Physik Department, Universität Siegen, Walter-Flex-Strasse 3, 57072 Siegen, Germany}
\affiliation[b]{School of Physics and Astronomy, University of Edinburgh, Edinburgh EH9 3JZ, UK}
\emailAdd{matthew.black@ed.ac.uk}
\emailAdd{martin.lang@uni-siegen.de}
\emailAdd{alexander.lenz@uni-siegen.de}
\emailAdd{zachary.wuethrich@uni-siegen.de}

\abstract{
    Theory predictions of heavy-hadron lifetime ratios critically depend on precise determinations of the dimension-six spectator effects arising from the double insertion of the weak effective $|\Delta B| = 1$ Hamiltonian.
    In the presence of beyond-standard-model (BSM) operators, the resulting $\Delta B = 0$ Hamiltonian features additional four-quark operators whose matrix elements need to be determined using non-perturbative methods.
    We present for the first time results for the non-perturbative hadronic matrix elements of the four-quark operators relevant for the description of the meson lifetime ratio $\tau\left(B^+\right) / \tau\left(B_d\right)$, obtained using heavy-quark effective theory (HQET) sum rules with the full BSM effective Hamiltonian.
    In addition, we recompute and update the bag parameters for the Standard Model operators.
}

\begin{document}
\maketitle

\section{Introduction}\label{sec:introduction}
 The Standard
Model of particle physics (SM) is an
extremely successful theory of the
microscopic world, but it leaves some
fundamental questions unanswered, such as the
origin of the matter-antimatter asymmetry
in the Universe or the nature of dark
matter. A promising route for identifying effects beyond the SM (BSM) are so-called 
indirect searches, where precision measurements for observables are compared with corresponding precise theory predictions. 
\\
In this work we will determine mandatory theory input for using lifetimes of heavy hadrons for indirect BSM searches.
\\
The lifetimes of the ground-state $B$ mesons with one heavy quark have been determined experimentally with a high precision and HFLAV \cite{HeavyFlavorAveragingGroupHFLAV:2024ctg} quotes\footnote{The HFLAV value $\tau\left(B_s\right) / \tau\left(B_d\right) = 1.0017 \pm 0.0034$ is derived from the individual lifetime values plus some additional measurements that determine only the lifetime ratio, leading to a slight shift of the ratio to a value larger than $1$.}
\begin{eqnarray}
\tau (B_d) = (1.517 \pm 0.004) \, 
\mbox{ps} \, ;
&&
\nonumber
\\
\tau (B^+) = (1.638 \pm 0.004) \, 
\mbox{ps} \, ;
&&
\frac{\tau (B^+)}{\tau (B_d)} = 1.076 \pm 0.004 \, ;
\nonumber
\\
\tau (B_s) = (1.516 \pm 0.006) \, 
\mbox{ps} \, ;
&&
\frac{\tau (B_s)}{\tau (B_d)} = 1.0017 \pm 0.0034 \, .
\end{eqnarray}
Lifetimes and lifetime ratios can be determined theoretically with the 
heavy-quark expansion (HQE)\footnote{See Refs.~\cite{Khoze:1983yp,Shifman:1984wx,Khoze:1986fa,Chay:1990da,Bigi:1991ir,Bigi:1992su,Bigi:1993fe,Blok:1993va,Mannel:1993su,Manohar:1993qn} for some early HQE references, 
Ref.~\cite{Lenz:2014jha} for a review of the historic development and Ref.~\cite{Albrecht:2024oyn} for a review of the state of the art.}.
The HQE provides a systematic framework to express total decay rates of hadrons containing  heavy (anti-)quarks $Q$ through a series of terms of increasing mass suppression as\footnote{In this work we are concerned with the lifetimes of the heavy-light $b$-flavoured mesons $B_q$, $q = u,d,s$, and will discuss the HQE with these mesons in mind.}
\begin{equation}
\Gamma(B_q) = 
\Gamma_3  +
\Gamma_5 \frac{\langle {\cal O}_5 \rangle}{m_b^2} + 
\Gamma_6 \frac{\langle {\cal O}_6 \rangle}{m_b^3} + \ldots  
 + 16 \pi^2 
\left( 
  \tilde{\Gamma}_6 \frac{\langle \tilde{\mathcal{O}}_6 \rangle}{m_b^3} 
+ \tilde{\Gamma}_7 \frac{\langle \tilde{\mathcal{O}}_7 \rangle}{m_b^4} + \ldots 
\right),
\label{eq:HQE}
\end{equation}
with perturbatively-calculable short-distance coefficients
\begin{equation}
\Gamma_i = \Gamma_i^{(0)} + \frac{\alpha_s}{\pi} \Gamma_i^{(1)} 
+ \left(\frac{\alpha_s}{\pi}\right)^2 \Gamma_i^{(2)} + \ldots \, ,  
\label{eq:HQE_pert}
\end{equation}
and non-perturbative matrix elements $\langle {\cal O}_i \rangle \equiv \langle {\cal O}_i \rangle_{B_q} \equiv \langle B_q | {\cal O}_i |B_q  \rangle/(2 M_{B_q})$  of  operators 
${\cal O}_i$ with mass-dimension $i$ and
no change of the $b$ quantum number, i.e.\ $\Delta B = 0$, and $M_{B_q}$ denoting the meson mass. 
The leading term in the HQE, $\Gamma_3$, describes the free $b$-quark decay and is free of non-perturbative effects. 
First power-suppressed corrections,  ${\cal O}_5$, arise  due to the kinetic and chromo-magnetic operators at order $1/m_b^2$.
Operators with mass-dimension six can originate from either two-quark operators, e.g.\ the Darwin operator, denoted by ${\cal O}_6$,
or from four-quark operators, denoted by ${\tilde{\cal O}}_6$.
The latter contributions originate from loop-enhanced diagrams, indicated by  the explicit factor of $16 \pi^2$ in \cref{eq:HQE}.
When considering lifetime ratios, the universal free $b$-quark decay cancels exactly and in the case of $\tau (B^+) /\tau (B_d)$ also all two-quark contributions cancel due to isospin symmetry of the matrix elements  $\langle {\cal O}_i \rangle$, while in the case
of $\tau (B_s) /\tau (B_d)$ two-quark contributions survive as $SU(3)_F$ breaking corrections, i.e.\
\begin{align}
\frac{\tau (B^+)}{\tau (B_d)}  = &\;
1 + \tau (B^+) \left\{ \vphantom{\int\limits_{-\infty}^{\infty}}
\hspace{3cm}
0
 \right.
\label{eq:B+overBd}
\\
&
\left.
 + 16 \pi^2 
\left[ \left( 
  \tilde{\Gamma}_6 \frac{\langle \tilde{\mathcal{O}}_6 \rangle}{m_b^3} 
+ \tilde{\Gamma}_7 \frac{\langle \tilde{\mathcal{O}}_7 \rangle}{m_b^4} + \ldots
\right)_{B_d}
\! \! \! \! \! \! -
\left( 
 \tilde{\Gamma}_6 \frac{\langle \tilde{\mathcal{O}}_6 \rangle}{m_b^3} + \tilde{\Gamma}_7 \frac{\langle \tilde{\mathcal{O}}_7 \rangle}{m_b^4} + \ldots
\right)_{B^+} \right]
\vphantom{\int\limits_{-\infty}^{\infty}} \right\}
 ,
\nonumber
\\
\frac{\tau (B_s)}{\tau (B_d)}  = &\; 
1 + \tau (B_s) \left\{ \vphantom{\int\limits_{-\infty}^{\infty}}
\Gamma_5 
\frac{
\langle {\cal O}_5 \rangle_{B_d}
-
\langle {\cal O}_5 \rangle_{B_s}
}{m_b^2} 
+ 
\Gamma_6 
\frac{
\langle {\cal O}_6 \rangle_{B_d}
 -
\langle {\cal O}_6 \rangle_{B_s}
}{m_b^3} + \ldots  
\right.
\label{eq:BsoverBd}
\\
&
\left.
 + 16 \pi^2 
\left[ \left( 
  \tilde{\Gamma}_6 \frac{\langle \tilde{\mathcal{O}}_6 \rangle}{m_b^3} 
+ \tilde{\Gamma}_7 \frac{\langle \tilde{\mathcal{O}}_7 \rangle}{m_b^4} + \ldots 
\right)_{B_d}
\! \! \! \! \! \! -
\left( 
  \tilde{\Gamma}_6 \frac{\langle \tilde{\mathcal{O}}_6 \rangle}{m_b^3} + \tilde{\Gamma}_7 \frac{\langle \tilde{\mathcal{O}}_7 \rangle}{m_b^4} + \ldots 
\right)_{B_s} \right]
\vphantom{\int\limits_{-\infty}^{\infty}} \right\}
 .
\nonumber
\end{align}
Hence lifetime ratios are
particularly sensitive to the contributions of the four-quark operators
$\tilde{\cal O}_i $. 
The Wilson coefficients of the dimension-six four-quark operators have been determined with the full charm-quark mass dependence in leading-order (LO) QCD, i.e.\ $\tilde{\Gamma}_6^{(0)}$, in 1996 in  Refs.~\cite{Uraltsev:1996ta,Neubert:1996we} and in next-to-leading-order (NLO) QCD, i.e.\ $\tilde{\Gamma}_6^{(1)}$,  in 2002 in
Refs.~\cite{Beneke:2002rj,Franco:2002fc}.
There were some exploratory studies of the matrix elements of the corresponding four-quark operators $\tilde{\mathcal{O}}_6$ within the framework of lattice QCD more than 20 years ago \cite{DiPierro:1998ty,DiPierro:1999tb,Becirevic:2001fy} and estimates within  the framework of QCD sum rules
\cite{Shifman:1978bx,Shifman:1978by} for the subdominant contributions of condensates to the lifetime matrix elements~\cite{Cheng:1998ia,Baek:1998vk}. 
The leading sum rule contribution to $\langle \tilde{\cal O}_6 \rangle$ arises as a perturbative three-loop contribution and these corrections were only available after the determination of the corresponding master integrals within the heavy-quark effective theory (HQET) in Ref.~\cite{Grozin:2008nu}.
Three-loop HQET sum rules were subsequently used to determine the four-quark matrix elements for 
$B_d$ mixing~\cite{Grozin:2016uqy,Kirk:2017juj,Kirk:2018kib},
$B_s$ mixing~\cite{King:2019lal,King:Thesis22},
$D$ mixing~\cite{Kirk:2017juj,Kirk:2018kib},
$B_d$ and $B^+$, as well as $D^0$ and 
$D^+$ lifetimes~\cite{Kirk:2017juj,Kirk:2018kib}
and $B_s$ and $D_s^+$ lifetimes~\cite{King:2021jsq,King:Thesis22}.
Recently two new different lattice studies have been ongoing for these quantities:
Ref.~\cite{Lin:2022fun} describes preliminary work towards the lifetime-difference operators for heavy mesons and baryons using position-space renormalisation in HQET, whereas Refs.~\cite{Black:2023vju,Black:Thesis24,Black:2024iwb} report progress towards determining both the lifetime-difference and the absolute-lifetime operators for heavy mesons using gradient flow~\cite{Narayanan:2006rf,Luscher:2010iy,Luscher:2013cpa} to renormalise the operators in full QCD. 
\\
The Wilson coefficients of the dimension-seven four-quark operators have been determined in LO QCD, i.e.\ $\tilde{\Gamma}_7^{(0)}$, in 1996 in  Ref.~\cite{Gabbiani:2004tp}, while the corresponding matrix elements have so far only been estimated in the vacuum insertion approximation (VIA).
\\
The current status of HQE predictions of the lifetime ratios within the SM~\cite{Albrecht:2024oyn, Lenz:2022rbq},
\begin{eqnarray}
\frac{\tau (B^+)}{\tau (B_d)} = 1.086 \pm 0.022 \, ,
&&
\frac{\tau (B_s)}{\tau (B_d)} = 1.003 \pm 0.006 \, ,
\end{eqnarray}
is in excellent agreement with the experimental measurements.
Theory predictions for the total decay
rates are typically not quoted (see Refs.~\cite{Albrecht:2024oyn, Lenz:2022rbq} for exceptions) due to large uncertainties originating from the $m_b^5$ dependence of $\Gamma_3$.
This will change, see Ref.~\cite{Egner:2024lay}, due to the recently-calculated next-to-next-to-leading 
order (NNLO) QCD corrections,
$\Gamma_3^{(2)}$, for the non-leptonic
decay channels \cite{Egner:2024azu}
and the NLO QCD corrections to the
chromomagnetic operator for the $b \to c \bar{u} d$ transition \cite{Mannel:2024uar}.
\\
BSM contributions to lifetime ratios arise either due to light new physics (see e.g.\ Ref.~\cite{Lenz:2024rwi})
 or due to a modification of
the effective $\abs{\Delta B} = 1$ Hamiltonian, describing e.g.\ non-leptonic tree-level decays.
The potential size of hypothetical
BSM effects in non-leptonic tree-level decays
was studied e.g.\ in Refs.~\cite{Bobeth:2014rda,Brod:2014bfa,Jager:2017gal,Lenz:2019lvd,Jager:2019bgk}.
Such effects could lead to a sizeable enhancement of the decay rate difference of neutral $B_d$ mesons \cite{Bobeth:2014rda}, $\Delta \Gamma_d$, which is not yet measured. 
It could further lead to substantial deviations in the  experimental extraction of the CKM angle $\gamma$ \cite{Brod:2014bfa} and could be the origin of the $B$ anomalies
\cite{Jager:2017gal,Jager:2019bgk} --- an ultraviolet (UV) completion of such a scenario was worked out in Ref.~\cite{Crivellin:2023saq}.
Recently this idea received some further interest due to the observation that 
for hadronic tree-level decays like $B_s \to D_s^- \pi^+$, the expectations from QCD factorisation \cite{Beneke:2000ry} differ significantly from experiment
\cite{Bordone:2020gao}; see Ref.~\cite{Huber:2016xod} for an earlier observation of this discrepancy.
Such a deviation could originate in underestimated corrections to QCD factorisation (see Refs.~\cite{Piscopo:2023opf,Beneke:2021jhp,Endo:2021ifc}) or in genuine
BSM effects; see e.g.\ Refs.~\cite{Cai:2021mlt,Iguro:2020ndk, Gershon:2021pnc,Fleischer:2021cct,Fleischer:2021cwb,Bordone:2021cca, Atkinson:2024hqp,Meiser:2024zea}.
\\
In order to study in a model-independent way
potential BSM contributions to non-leptonic decays, the usual effective $\abs{\Delta B}= 1 $ Hamiltonian (see e.g.\ the review \cite{Buchalla:1995vs}) can be extended to 20 BSM operators; cf.\ Refs.~\cite{Jager:2017gal,Jager:2019bgk,Cai:2021mlt}.
With this new set of operators, the contributions to mixing and lifetimes ratios have been determined in LO QCD,
i.e.\ $\tilde{\Gamma}_6^{(0),  \text{BSM}}$,
for the decay $b \to c \bar{c}s $
\cite{Jager:2017gal,Jager:2019bgk}
and the decay $b \to c \bar{u} d $
\cite{Lenz:2022pgw}.\footnote{Currently the determination of 
$\Gamma_3^{(0), \text{BSM}}$, 
$\Gamma_5^{(0), \text{BSM}}$ 
and
$\Gamma_6^{(0), \text{BSM}}$
for these new $\abs{\Delta B}= 1 $ operators is in progress \cite{DarwinBSM}. Expressions for $\Gamma_3^{(0), \text{BSM}}$ for the $
b \to c \bar{u} d$ channel have been presented in Ref.~\cite{Meiser:2024zea}.
}
For the case of lifetimes this leads also to new $\Delta B = 0$ four-quark operators, whose matrix elements will be determined for the first time in this work with HQET sum rules. In that respect we also recalculated the matrix elements for the SM case and found a typo in the original work 
\cite{Kirk:2017juj,King:2021jsq}. Updated expressions and numerical values will be given in this work.

The remainder of this paper is organised as follows.
In \cref{sec:effHam} we introduce the effective operators of the general $\Delta B = 0$ Hamiltonian within HQET, before setting up the sum rule for the object of interest in \cref{sec:sumrule}. 
\Cref{sec:calculation} is dedicated to a discussion of some aspects of the perturbative calculation, \cref{sec:condensates} contains some details about the calculation of the condensate contributions, and then results are presented in \cref{sec:results}.
Finally, we conclude in \cref{sec:conclusion}.
\section{\texorpdfstring{$\mathbf{\Delta B = 0}$}{Delta B = 0} effective four-fermion operators}\label{sec:effHam}
Generic effects of BSM particles interacting at some high-energy UV scale will affect low-energy observables through effective operators once
they have been integrated out.
The resulting effective $\abs{\Delta B} = 1$ Hamiltonian (see e.g.\ Refs.~\cite{Buchalla:1995vs,Buras:2000if,Cai:2021mlt,Jager:2019bgk,Lenz:2022pgw})
\begin{equation}
    \mathcal{H}_{\mathrm{eff}}^{\abs{\Delta B} = 1} = \frac{4 G_F}{\sqrt{2}} V_{q_1 b}^* V_{q_2 q_3} \sum_{i} C_i^{(\prime)} \mathcal{Q}_i^{(\prime)}  + \mathrm{h.c.}
\end{equation}
is parameterised in terms of the operators $\mathcal{Q}_i^\prime$ and the Wilson coefficients $C_i^{(\prime)}$ which we leave unspecified in absence of a specific model under consideration.
Using the optical theorem, the total decay width of a meson $B_q$ containing a heavy bottom quark and a light quark $q$ is given by
\begin{equation}
    \Gamma \left(B_q \right) = \frac{1}{2 M_{B_q}} \mathrm{Im} \left[ \mel**{B_q}{\,\mathrm{i} \int \mathrm{d}^4 x \, \mathbf{T} \left\{ \mathcal{H}_{\mathrm{eff}}^{\abs{\Delta B} = 1} \left(x\right) \mathcal{H}_{\mathrm{eff}}^{\abs{\Delta B} = 1} \left(0\right) \right\} }{B_q} \right] \,,
\end{equation}
where $\mathbf{T}$ denotes the time-ordering symbol and $M_{B_q}$ is the mass of the unstable meson.
The subsequent operator-product expansion (OPE), known as the HQE, gives rise at the scale $\mu \sim m_b$ to a series of operators as laid out in \cref{sec:introduction}, among which four-quark $\Delta B = 0$ operators $\bar{b} \Gamma q \, \bar{q} \Gamma^\prime b$ appear, suppressed by $\left(\Lambda_{\mathrm{QCD}} / m_b\right)^3$ with respect to the leading term of the HQE \cite{Lenz:2014jha}.
The Wilson coefficients of these operators are sensitive to the flavour of the
light quark $q$ and these contributions are therefore referred to as 
\emph{spectator effects}.
These play a dominant role in the description of lifetime ratios of different bottom hadrons.
In the following we set up the basis of operators considered in this paper.

\subsection{Physical operators in QCD}\label{subsec:physopsQCD}
A double insertion of the most general BSM $\abs{\Delta B} = 1$ Hamiltonian induces at mass-dimension six a set of $20$ physical $\Delta B = 0$ operators in QCD --- denoted generically by $\tilde{\cal O}_6$ in \cref{eq:HQE}.
These are
\begin{equation}\label{eq:QCDphysops}
\begin{aligned}
   O_1^q \equiv Q_1^q &\equiv \bar{b} \gamma_\mu \left( 1- \gamma_5\right) q \, \bar{q} \gamma^\mu \left(1 - \gamma_5\right) b \,, \\
   O_2^q \equiv Q_2^q &\equiv \bar{b} \left( 1 - \gamma_5\right) q \, \bar{q} \left(1 + \gamma_5\right) b \,, \\ 
   O_3^q \equiv T_1^q &\equiv \bar{b} \gamma_\mu \left( 1 - \gamma_5\right) T^a q \, \bar{q} \gamma^\mu \left( 1 - \gamma_5\right) T^a b \,, \\
   O_4^q \equiv T_2^q &\equiv \bar{b} \left( 1 - \gamma_5 \right) T^a q \, \bar{q} \left( 1 + \gamma_5\right) T^a b \,, \\
   O_5^q \equiv Q_3 &\equiv \bar{b} \gamma_\mu \left(1 - \gamma_5\right) q \, \bar{q} \gamma^\mu \left(1 + \gamma_5\right) b \,, \\
   O_6^q \equiv Q_4 &\equiv \bar{b} \left(1 - \gamma_5\right) q \, \bar{q} \left(1 - \gamma_5\right) b \,, \\
   O_7^q \equiv T_3 &\equiv \bar{b} \gamma_\mu \left(1 - \gamma_5\right) T^a q \, \bar{q} \gamma^\mu \left(1 + \gamma_5\right) T^a b \,, \\
   O_8^q \equiv T_4 &\equiv \bar{b} \left(1 - \gamma_5\right) T^a q \, \bar{q} \left(1 - \gamma_5\right) T^a b \,, \\
   O_9^q \equiv Q_5 &\equiv \bar{b} \sigma_{\mu\nu} \left(1 - \gamma_5\right) q \, \bar{q} \sigma^{\mu\nu} \left(1 - \gamma_5\right) b \,, \\
   O_{10}^q \equiv T_5 &\equiv \bar{b} \sigma_{\mu\nu} \left(1 - \gamma_5\right) T^a q \, \bar{q} \sigma^{\mu\nu} \left(1 - \gamma_5\right) T^a b \,, \\
   O_i^{\prime \, q} & \equiv \left. O_i^q\right|_{1 \mp \gamma_5 \to 1 \pm \gamma_5} \,, \qquad i = 1, \ldots, 10 \,.
\end{aligned}
\end{equation}
The operator labels ${Q}$ and ${T}$ have been used historically for the colour-singlet and colour-octet operators, and we use them or ${O}$ interchangeably in order to make equations as compact as possible.
In the SM only the four operators 
$\left\{Q_1^q, Q_2^q, T_1^q, T_2^q\right\}$
arise.

\subsection{Lifetime ratios}\label{subsec:lifetimeratiosQCD}
Within this paper our focus will be the lifetime ratio $\tau \left(B^+\right) / \tau\left(B_d\right) = \Gamma\left(B_d\right) / \Gamma\left(B^+\right)$, 
where the expression of \cref{eq:B+overBd} can be further specified in order to
explicitly distinguish the contributions of different spectator quarks $q$ as
\begin{equation}
\label{eq:B+overBd_2}
    \frac{\Gamma\left(B_d\right)}{\Gamma\left(B^+\right)} 
    = 1 + \tau (B^+)
    \frac{16 \pi^2}{m_b^3} \sum_{q=u,d,s,c} \sum_{i}\tilde{\Gamma}_{6,i}^q \left( 
    \frac{\overline{\expval{O^q_i}}_{B_d}}{2M_{B_d}} - \frac{\overline{\expval{O^q_i}}_{B^+}}{2M_{B^+}} \right) + \mathcal{O} \left(\frac{1}{m_b^4}\right) \,,
\end{equation}
with a slightly modified notation for the matrix elements of the four-quark operators 
\begin{equation}
    \overline{\expval{O_i^{q^\prime}}}_{B_q} \equiv \mel**{B_q}{O_i^{q^\prime}}{B_q} \,.
\end{equation}
The different Dirac and colour structures are denoted by the index $i$.
As expressed in \cref{eq:B+overBd}, contributions of two-quark operators (mass-dimensions three, five, six, ...) cancel 
to an excellent precision due to isospin symmetry in the difference of total decay rates.
The coefficients  $\tilde{\Gamma}_6^q$ denote all short-distance contributions at mass-dimension six obtained from a double insertion of the $\abs{\Delta B} = 1$ Hamiltonian into the spectator-type diagrams.
\\
Using the isospin relations
\begin{equation}
   \frac{\overline{\expval{O_{i}^{s}}}_{B_d}}{2M_{B_d}} = 
   \frac{\overline{\expval{O_{i}^{s}}}_{B^+}}{2M_{B^+}}
   \,, 
   \qquad 
   \frac{\overline{\expval{O_{i}^{c}}}_{B_d}}{2M_{B_d}}
   = \frac{\overline{\expval{O_{i}^{c}}}_{B^+}}{2M_{B^+}}\,,
\end{equation}
and
\begin{equation}
    \frac{\overline{\expval{O_{i}^{u}}}_{B_d}}{2M_{B_d}}
    =\frac{\overline{\expval{O_{i}^{d}}}_{B^+}}{2M_{B^+}}
    \,, \qquad 
    \frac{\overline{\expval{O_{i}^{d}}}_{B_d}}{2M_{B_d}}
    = \frac{\overline{\expval{O_{i}^{u}}}_{B^+}}{2M_{B^+}} \,,
\end{equation}
we find that the contributions due to strange and charm spectator quarks as well as all contributions due to eye-contractions cancel exactly. 
We are left with
\begin{equation}
\label{eq:B+overBd_3}
    \frac{\Gamma\left(B_d\right)}{\Gamma\left(B^+\right)} 
    = 1 + \tau (B^+) \frac{16 \pi^2}{m_b^3}
    \sum_{i}
    \left(    \tilde{\Gamma}_{6,i}^d -   \tilde{\Gamma}_{6,i}^u 
    \right)
    \frac{\overline{\expval{ O^u_i - O^d_i}}_{B^+}}{2M_{B^+}}
    + \mathcal{O} \left(\frac{1}{m_b^4}\right) \, .
\end{equation}
Therefore, we introduce the isospin-breaking combinations of operators
\begin{equation}
    \label{eq:IBopsQCD}
    O_i \equiv O_i^u - O_i^d \,,
\end{equation}
whose matrix elements are parameterised in terms of the leptonic decay constants and \emph{bag parameters} as
\begin{align}\label{eq:definitionAiQCD}
    \overline{\expval{Q_i}}_{B^+} \left(\mu\right) &= A_i f_{B^+}^2 M_{B^+}^2 B_i \left(\mu\right) \,, \\
    \overline{\expval{T_i}}_{B^+} \left(\mu\right) &= A_i f_{B^+}^2 M_{B^+}^2 \epsilon_i \left(\mu\right) \,,
\end{align}
where
\begin{equation}\label{eq:afactorsqcd}
    A_1 = 1 \,, \qquad A_2 = \frac{M_{B^+}^2}{\left(m_b^{\mathrm{OS}} + m_u\right)^2} \,.
\end{equation}
We will neglect $m_u$ in the following.
The definition of the bag parameters is inspired by the \emph{vacuum insertion (saturation) approximation} (VIA), which presumes that the matrix element of a four-quark operator factorises into the product of two-quark matrix elements,
\begin{equation}
    \langle B_q|\bar{b}\Gamma q\,\bar{q}\Gamma^\prime b|B_q\rangle \stackrel{\text{VIA}}{\equiv} \langle B_q|\bar{b}\Gamma q|0\rangle\,\langle 0|\bar{q}\Gamma^\prime b|B_q\rangle.
\end{equation}
Therefore in the VIA, the bag parameters read
\begin{align}
    B_i \left(\mu\right) &= 1 \,, &
    \epsilon_i \left(\mu\right) = 0 \,,
\end{align}
and in general their deviations from these values describe the violation of the VIA in these four-quark matrix elements.
It is expected that this deviation is not large, however calculating its size is important for accurately and precisely predicting quantities such as the lifetime ratios of interest in this work.

The contributions from the isospin-breaking operators discussed here are the only ones relevant for the lifetime ratio $\tau \left(B^+\right) / \tau \left(B_d\right)$.
They arise solely from the diagrams in which the four-quark operator $O_i^{q^\prime}$ connects to both valence quark lines $\bar{b},\, q$ and therefore must have $q = q^\prime$.
In these diagrams one has $\overline{\expval{O_i^u - O_i^d}}_{B^+} = \overline{\expval{O_i^u}}_{B^+}$.
The second class of diagrams in which the $q^\prime$ quark line of the effective operator is contracted with itself (the so-called `eye' diagrams), gives the same contribution for all $q$ up to tiny $SU(3)_F$-breaking corrections and is therefore irrelevant for decay rate differences; these diagrams can have $q = q^\prime$ or $q \neq q^\prime$.
For the matrix elements relevant to absolute meson lifetimes, these additional contributions have to be added, i.e.\ for the matrix elements relevant in absolute lifetimes the contributions from all eye diagrams with $q = q^\prime$ and $q \neq q^\prime$ have to be added on top of the matrix elements we determine in this work.
For the absolute lifetime $\tau(B_s)$ or the lifetime ratio $\tau(B_s)/\tau(B_d)$, additional strange-quark mass corrections will arise as $SU(3)_F$-breaking effects; see Refs.~\cite{King:2021jsq,King:Thesis22}.

\subsection{Physical operators and lifetime ratio in HQET}\label{subsec:physopsHQET}
In HQET, the $b$ quark field will be replaced by the heavy quark field $h$ with mass $m_Q$.
We find  four SM operators
\begin{equation}
\begin{aligned}
   \tilde{O}_1^q \equiv \tilde{Q}_1^q &\equiv \bar{h} \gamma_\mu \left( 1- \gamma_5\right) q \, \bar{q} \gamma^\mu \left(1 - \gamma_5\right) h \,, \\
   \tilde{O}_2^q \equiv \tilde{Q}_2^q &\equiv \bar{h} \left( 1 - \gamma_5\right) q \, \bar{q} \left(1 + \gamma_5\right) h \,, \\ 
   \tilde{O}_3^q \equiv \tilde{T}_1^q &\equiv \bar{h} \gamma_\mu \left( 1 - \gamma_5\right) T^a q \, \bar{q} \gamma^\mu \left( 1 - \gamma_5\right) T^a h \,, \\
   \tilde{O}_4^q \equiv \tilde{T}_2^q &\equiv \bar{h} \left( 1 - \gamma_5 \right) T^a q \, \bar{q} \left( 1 + \gamma_5\right) T^a h \,,
\end{aligned}
\end{equation}
supplemented by twelve BSM operators
\begin{equation}
\begin{aligned}
   \tilde{O}_5^q \equiv \tilde{Q}_3^q &\equiv \bar{h} \gamma_\mu \left(1 - \gamma_5\right) q \, \bar{q} \gamma^\mu \left(1 + \gamma_5\right) h \,, \\
   \tilde{O}_6^q  \equiv \tilde{Q}_4^q &\equiv \bar{h} \left(1 - \gamma_5\right) q \, \bar{q} \left(1 - \gamma_5\right) h \,, \\
   \tilde{O}_7^q \equiv \tilde{T}_{3}^q &\equiv \bar{h} \gamma_\mu \left(1 - \gamma_5\right) T^a q \, \bar{q} \gamma^\mu \left(1 + \gamma_5\right) T^a h \,, \\
   \tilde{O}_8^q \equiv \tilde{T}_{4}^q & \equiv \bar{h} \left(1 - \gamma_5\right) T^a q \, \bar{q} \left(1 - \gamma_5\right) T^a h \,, \\
   \tilde{O}_i^{\prime \, q} & \equiv \left. \tilde{O}_i^q\right|_{1 \mp \gamma_5 \leftrightarrow 1 \pm \gamma_5} \,, \qquad i = 1, \ldots, 8 \,.
\end{aligned}
\end{equation}
The  four additional operators $\bar{h} \sigma_{\mu\nu} P_\alpha (T^a) q \, \bar{q} \sigma^{\mu\nu} P_\alpha (T^a) h$, with $P_\alpha = 1 \mp \gamma_5$, that appear in the QCD case can be reduced to linear combinations of the operators $\tilde{O}_{5-8}^{(\prime)q}$ \cite{Lenz:2022pgw}, such that the operators involving $\sigma_{\mu\nu}$ can be expressed as
\begin{equation}\label{eq:eomtensorHQET}
    \bar{h}\sigma_{\mu\nu}P_{L,R}q \, \bar{q}\sigma^{\mu\nu} P_{L,R}h = -4 \left[ \bar{h}P_{L,R}q \, \bar{q}P_{L,R} h - \bar{h}\gamma_\mu P_{L,R} q \, \bar{q} \gamma^\mu P_{R,L} h\right] + \mathcal{O} \left(\frac{1}{m_Q}\right) \,.
\end{equation}
By using short-distance Wilson coefficients calculated in HQET \cite{Franco:2002fc} rather than QCD as well as the HQET $\Delta B = 0$ operators in \cref{eq:B+overBd_3}, one can express the decay rate in terms of isospin-breaking combinations of the HQET operators,
\begin{equation}
    \label{eq:IBopsHQET}
    \tilde{O}_i \equiv \tilde{O}_i^u - \tilde{O}_i^d \,,
\end{equation}
whose matrix elements we parameterise in terms of bag parameters and the HQET meson decay constant $F(\mu)$,
\begin{alignat}{3}
       &\overline{\expval{\tilde{Q}_i}}_{B^+} \left(\mu\right) &&= \tilde{A}_i F^2 \left(\mu\right) \tilde{B}_i \left(\mu\right) \,, \\
       &\overline{\expval{\tilde{T}_i}}_{B^+} \left(\mu\right) &&= \tilde{A}_i F^2 \left(\mu\right) \tilde{\epsilon}_i \left(\mu\right) \,,
\end{alignat}
with $\tilde{A}_{1,2} = +1$ and $\tilde{A}_{3,4} = -1$ in HQET.
The HQET bag parameters $\tilde{B}_i \left(\mu\right)$, $\tilde{\epsilon}_i \left(\mu\right)$ are the main result of this paper.
In the VIA, they read
\begin{align}
    \tilde{B}_i \left(\mu\right) &= 1 \,, &
    \tilde{\epsilon}_i \left(\mu\right) = 0 \,.
\end{align}

\subsection{Evanescent operators}
Higher-order calculations involving four-fermion operators, if performed in dimensional regularisation with $d = 4 - 2 \epsilon$, usually require the introduction of additional evanescent operators that vanish once the limit $d \to 4$ is taken, but contribute to the matrix elements of physical operators in the matching.
For our calculation we need $6+6$ evanescent operators defined in terms of HQET fields\footnote{Note that in QCD two additional evanescent operators have to be introduced; the basis of evanescent operators in QCD is listed in \cref{sec:evanQCD}.},
\begin{equation}\label{eq:definitionevanescent}
    \begin{aligned}
       \tilde{E}_{1}^q &\equiv \bar{h} \gamma_{\mu\nu\rho} \left(1 - \gamma_5\right) q \, \bar{q} \gamma^{\rho\nu\mu} \left(1 - \gamma_5\right) h \quad - \left(4 + a_1 \epsilon\right) \tilde{O}_{1}^q \,, \\
       \tilde{E}_{2}^q &\equiv \bar{h} \gamma_{\mu\nu} \left(1 - \gamma_5\right) q \, \bar{q} \gamma^{\nu\mu} \left(1 + \gamma_5\right) h \quad - \left(4 + a_2 \epsilon\right) \tilde{O}_{2}^q \,, \\
       \tilde{E}_{3}^q &\equiv \bar{h} \gamma_{\mu\nu\rho} \left(1 - \gamma_5\right) T^a q \, \bar{q} \gamma^{\rho\nu\mu} \left(1 - \gamma_5\right) T^a h \quad - \left(4 + a_1 \epsilon\right) \tilde{O}_{3}^q \,, \\
       \tilde{E}_{4}^q &\equiv \bar{h} \gamma_{\mu\nu} \left(1 - \gamma_5\right) T^a q \, \bar{q} \gamma^{\nu\mu} \left(1 + \gamma_5\right) T^a h \quad - \left(4 + a_2 \epsilon\right) \tilde{O}_{4}^q \,, \\
       \tilde{E}_{5}^q &\equiv \bar{h} \gamma_{\mu\nu\rho} \left(1 - \gamma_5\right) q \, \bar{q} \gamma^{\rho\nu\mu} \left(1 + \gamma_5\right) h \quad - \left(16 + a_3 \epsilon \right) \tilde{O}_5^q \,, \\
       \tilde{E}_{6}^q &\equiv \bar{h} \gamma_{\mu\nu\rho} \left(1 - \gamma_5\right) T^a q \, \bar{q} \gamma^{\rho\nu\mu} \left(1 + \gamma_5\right) T^a h \quad - \left(16 + a_3 \epsilon \right) \tilde{O}_7^q \,, \\
    \end{aligned}
\end{equation}
plus the corresponding primed operators, with $\gamma^{\mu_1\cdots\mu_n} = \gamma^{\mu_1} \cdots \gamma^{\mu_n}$.
The fact that \cref{eq:eomtensorHQET} holds in $d$ dimensions implies that no evanescent operators correcting for $\tilde{O}_{6,8}^{(\prime)q}$ have to be introduced for this calculation.
In writing \cref{eq:definitionevanescent} we have introduced the parameters $a_{1,2,3}$ in the definition of the evanescent operators.
These parameters amount to redefinitions of the evanescent operators by terms of $\mathcal{O} \left( \epsilon\right)$ and must therefore drop out of any physical observable consistently calculated at NLO.
However, as the NLO anomalous dimension matrix for the HQET $\Delta B = 0$ operators has so far not been determined, a full independence of the scheme of evanescent operators is not achieved and we decide to keep these parameters, varying them in our numerical analysis as an estimate of uncertainty.

For the discussion of lifetime ratios we again define the isospin-breaking combinations
\begin{equation}
    \tilde{E}_i \equiv \tilde{E}_i^u - \tilde{E}_i^d \,.
\end{equation}

\section{The sum rule}\label{sec:sumrule}
Our starting point for the HQET sum rule is the three-point correlator~\cite{Grozin:2016uqy,Kirk:2017juj}
\begin{equation}\label{eq:correlatorintro}
    K_{\tilde{O}_i}\left(\omega_1, \omega_2\right) = \int \mathrm{d}^d x_1 \mathrm{d}^d x_2 \mathrm{e}^{\mathrm{i} \left(p_1 \cdot x_1 - p_2 \cdot x_2\right)} \mel**{0}{\mathbf{T} \left\{ \left[ \tilde{j} \left(x_2\right) \tilde{O}_i\left(0\right) \tilde{j}^\dagger \left(x_1\right) \right] \right\} }{0} \,,
\end{equation}
with the residual energies $\omega_{1,2} = p_{1,2} \cdot v$ and the interpolating current
\begin{equation}
    \tilde{j} = \bar{q} \gamma_5 h \,.
\end{equation}
For large and negative $\omega_{1,2}$, the correlator is highly virtual and can be computed in the OPE picture as a series of perturbative contributions and vacuum condensates.
To relate this correlator to the continuous hadronic spectrum in a sum rule, we can translate from the highly-virtual regime to the positive, real axis via the dispersion relation
\begin{equation}
    K_{\tilde{O}_i}(\omega_1,\omega_2) = \int_0^\infty \mathrm{d}\nu_1\,\mathrm{d}\nu_2\,\frac{\rho_{\tilde{O}_i}(\nu_1,\nu_2)}{(\nu_1-\omega_1)(\nu_2-\omega_2)} + [\text{subtraction terms}],
\end{equation}
where $\rho_{\tilde{O}_i}$ is the \emph{spectral density} of the correlator. 
When the correlator is analytic in the complex $\omega_{1,2}$ planes except for discontinuities along the positive, real axis, then the spectral density is defined as the double discontinuity of the correlator. 
Some important details on taking the double discontinuity of these correlators are given in \cref{sec:DoubleDisc}.

\begin{figure}[t]
    \centering
    \includegraphics[width=0.5\textwidth]{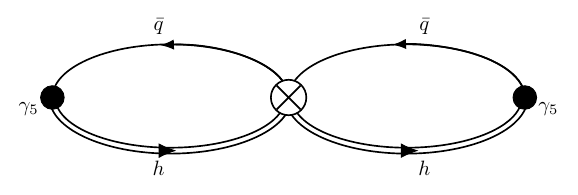}
    \caption{\label{fig:LO_FD} Leading-order Feynman diagram of the three-point correlation function defined in \cref{eq:correlatorintro}.}
\end{figure}

The leading Feynman diagram for the correlator in \cref{eq:correlatorintro} arises at two-loop order and is shown in \cref{fig:LO_FD}.
However, it is possible to split this diagram by a vertical cut through the effective operator, such that no momentum or state with non-vacuum quantum numbers crosses the cut.
This diagram is therefore factorisable in both QCD/HQET and contributes to the decay constant, but not to the deviation of the bag parameters from the VIA.
In fact, all diagrams in the OPE picture can be categorised into factorisable and non-factorisable contributions. 
The factorisable pieces obey the VIA and decompose to the product of two two-point correlators $\Pi(\omega)$, while the non-factorisable pieces describe the deviation from the VIA, i.e.~the bag parameters.
Therefore the spectral density can be written as
\begin{equation}
    \rho_{\tilde{O}_i}(\nu_1,\nu_2) = A_{\tilde{O}_i}\rho_{\Pi}(\nu_1)\rho_{\Pi}(\nu_2) + \Delta\rho_{\tilde{O}_i}(\nu_1,\nu_2).
\end{equation}
In this work, we are solely concerned with calculating $\Delta\rho_{\tilde{O}}$, while we take the result for $\rho_\Pi$ at two-loop order~\cite{Broadhurst:1991fc,Bagan:1991sg,Neubert:1991sp}.

Using the decomposition of the spectral density and a double Borel transformation of the sum rule\footnote{For further details, see for instance Refs.~\cite{Grozin:2016uqy,Kirk:2017juj,King:Thesis22}.}, we can isolate the non-factorisable contributions and derive a sum rule for the deviations from the VIA $\Delta \tilde{B}_{\tilde{O}_i}=\tilde{B}_{\tilde{O}_i}-\tilde{B}_{\tilde{O}_i}^{\text{VIA}}$, reading\footnote{The continuum cutoff $\omega_c$ will be discussed further in \cref{subsec:condensates}.}
\begin{equation}\label{eq:tradSR}
    \Delta \tilde{B}_{\tilde{O}_i} = \frac{1}{\tilde{A}_{\tilde{O}_i}}\frac{\displaystyle{\int_0^{\omega_c}\mathrm{d}\nu_1\,\mathrm{d}\nu_2\,e^{-\frac{\nu_1}{t_1}-\frac{\nu_2}{t_2}}\Delta\rho_{\tilde{O}_i}}(\nu_1,\nu_2)}{\left[\displaystyle{\int_0^{\omega_c}\mathrm{d}\nu_1\,e^{-\frac{\nu_1}{t_1}}\rho_\Pi(\nu_1)} \right]\,\left[\displaystyle{\int_0^{\omega_c}\mathrm{d}\nu_2\,e^{-\frac{\nu_2}{t_2}}\rho_\Pi(\nu_2)} \right]}.
\end{equation}
Following Ref.~\cite{Kirk:2017juj}, for the perturbative contribution to the spectral density we can introduce an arbitrary weight function which allows us to remove the integration from the sum rule entirely, leading to a simple expression for the bag parameter as
\begin{equation}\label{eq:DeltaBag}
    \Delta \tilde{B}_{\tilde{O}_i}^{\text{pert}}(\mu_\rho) = \frac{C_F}{N_c \tilde{A}_{\tilde{O}_i}}\frac{\alpha_s(\mu_\rho)}{4\pi}\;r_{\tilde{O}_i}\left(1,\log\frac{\mu_\rho^2}{4\bar{\Lambda}^2}\right),
\end{equation}
where the $r$-functions are defined via the spectral densities through
\begin{equation}
    \Delta \rho_{\tilde{O}_i} (\mu_\rho) = \frac{N_cC_F}{4}\frac{\nu_1^2\nu_2^2}{\pi^4}\frac{\alpha_s (\mu_\rho)}{4\pi}\,r_{\tilde{O}_i}(x,L_\nu).
\end{equation}

\section{Details of the perturbative calculation}\label{sec:calculation}
\begin{figure}[t]
    \centering
    \includegraphics[width=\textwidth]{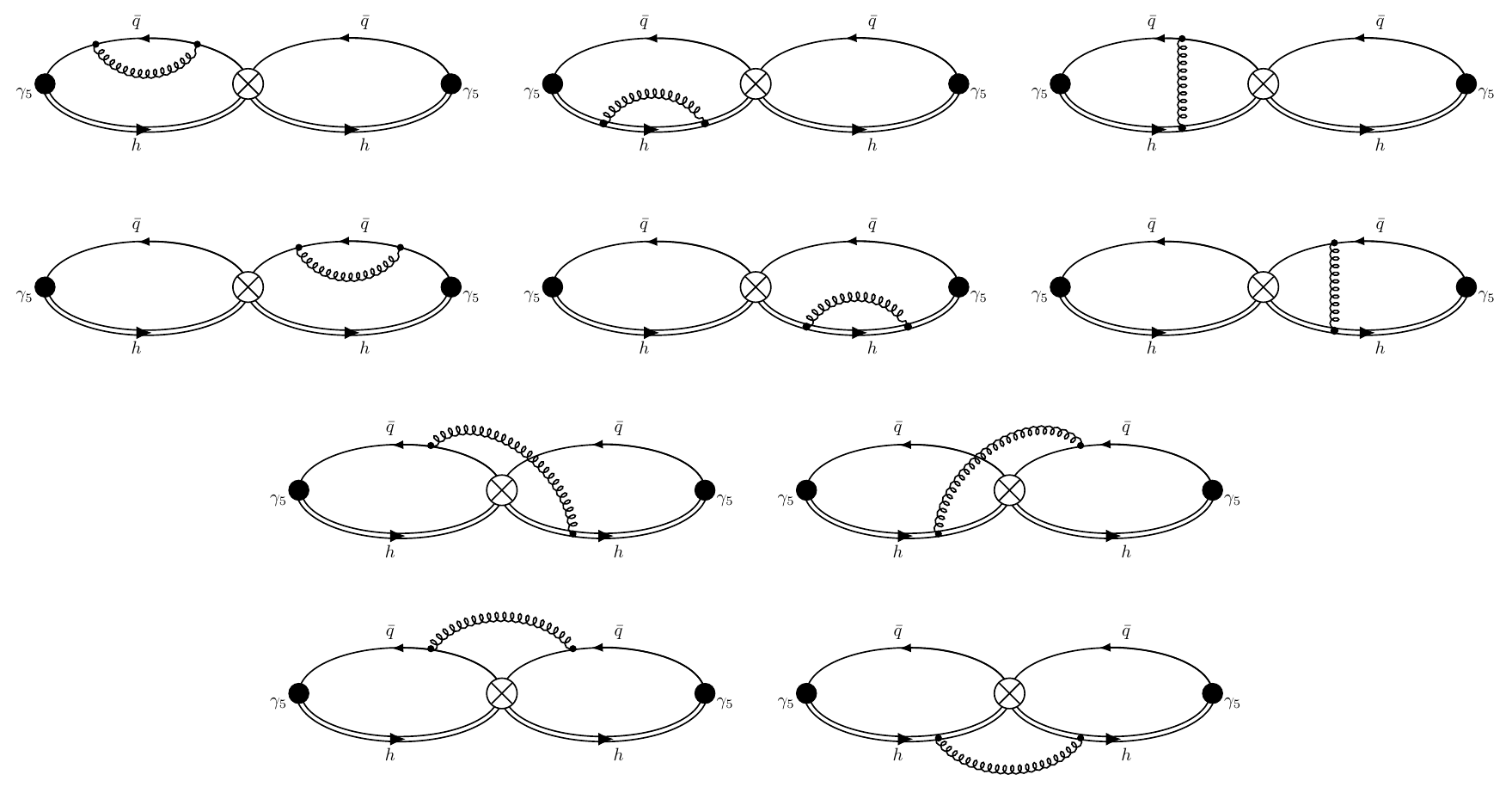}
    \caption{\label{fig:NLO_FD} Next-to-leading-order Feynman diagrams of the three-point correlation function defined in \cref{eq:correlatorintro}. The non-factorisable diagrams are those in the last two rows where the gluon line crosses the operator insertion.}
\end{figure}
\begin{figure}[t]
    \centering
    \includegraphics[width=0.5\textwidth]{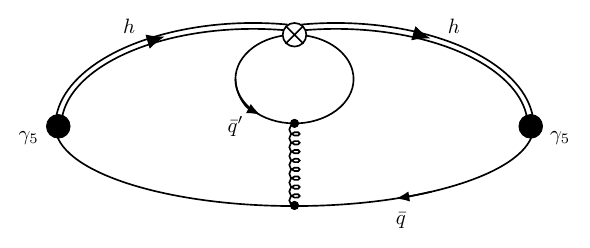}
    \caption{\label{fig:Eye_FD} Example of a next-to-leading order `eye' Feynman diagram of the three-point correlation function defined in \cref{eq:correlatorintro}. This contribution cancels when considering the isospin-breaking operators.}
\end{figure}

At NLO in QCD, additional diagrams beyond \cref{fig:LO_FD} appear with a gluon exchange between fermion lines.
These diagrams are shown in \cref{fig:NLO_FD}, including the first non-factorisable diagrams which contribute to the bag parameters.
When considering absolute lifetimes, there are also additional `eye' diagrams (see e.g.~\cref{fig:Eye_FD}), however these cancel when considering the isospin-breaking combinations introduced in \cref{eq:IBopsHQET}.

To compute the corrections to the bag parameters, we have to evaluate the non-factorisable three-loop Feynman diagrams with insertions of all possible physical HQET operators.
However, the insertions of $\tilde{O}_{1,2,5,6}$ vanish due to their colour structure such that we are left with only contributions to $\tilde{\epsilon}_{3,4,7,8}$.

Since the diagrams to be evaluated are of three-loop order, they necessitate some degree of automatisation in solving them.
We therefore compute the renormalised three-loop correlator using two different setups.
In the first approach, Feynman diagrams were generated with \texttt{qgraf} \cite{Nogueira:1991ex} and then written into symbolic expressions and manipulated using \texttt{tapir} \cite{Gerlach:2022qnc} and \texttt{FORM} \cite{Vermaseren:2000nd,Kuipers:2012rf,Ruijl:2017dtg} routines.
Due to the nature of the HQET propagators involving the residual energy, the partial fraction decomposition as a preparatory step for the integration-by-parts (IBP) reduction was performed with the help of \texttt{Mathematica}.
Dirac traces were computed in the Breitenlohner-Maison-'t Hooft-Veltman (BMHV) scheme using \texttt{TRACER} \cite{Jamin:1991dp}.
Finally, using IBP relations \cite{Chetyrkin:1981qh,Laporta:2000dsw} the Feynman integrals were reduced to a set of $11$ master integrals using \texttt{Kira} \cite{Maierhofer:2017gsa,Klappert:2020nbg}, listed in \cref{sec:masterintegrals}.

In the second approach, the non-factorisable Feynman diagrams were determined by hand and the symbolic expressions were manipulated using \texttt{Mathematica} where the program \texttt{FeynCalc}~\cite{Mertig:1990an,Shtabovenko:2016sxi,Shtabovenko:2023idz} was used to compute the Dirac traces in the BMHV scheme. The IBP reduction was then also performed using \texttt{Mathematica} in combination with \texttt{FeynCalc}, \texttt{LiteRed} \cite{Lee:2012cn,Lee:2013mka}, and \texttt{FIRE} \cite{Smirnov:2008iw,Smirnov:2019qkx}.
We have verified that the bare correlators agree exactly in both approaches.

The renormalisation of the NLO correlators involves mixing amongst the physical operators and also mixing of evanescent operators into physical operators.
In order to renormalise the three-loop bare correlator, one has to compute the two-loop factorisable diagrams with insertions of the physical and evanescent operators.
Due to their colour structure, only the insertions of colour-singlet operators yield non-vanishing contributions.
We can express the renormalised quantity as~\cite{Kirk:2017juj}
\begin{equation}
    K_{\tilde{O}_i}^{(1)} = K_{\tilde{O}_i}^{(1),\mathrm{bare}} + \frac{\alpha_s}{4\pi} \frac{1}{2\epsilon} \left[ \left(\hat{\tilde{\gamma}}_{\tilde{O}_i \tilde{O}_j}^{(0)} - 2 \hat{\tilde{\gamma}}_{\tilde{j}}^{(0)} \delta_{ij} \right) K_{\tilde{O}_j}^{(0)} + \hat{\tilde{\gamma}}_{\tilde{O}_i \tilde{E}_j}^{(0)} K_{\tilde{E}_j}^{(0)} \right] \,,
\end{equation}
where $\hat{\tilde{\gamma}}_{\tilde{j}}^{(0)} = - 3 C_F$ and $\tilde{O}_i$ denotes any SM or BSM operator, and summation over $j$ is implied. 
The anomalous dimension matrices $\hat{\tilde{\gamma}}_{\tilde{O}\tilde{O}}^{(0)}$ and $\hat{\tilde{\gamma}}_{\tilde{O}\tilde{E}}^{(0)}$ are listed in \cref{sec:adms}.

Finally, the master integrals have already been computed in Refs.~\cite{Grozin:2008nu,Grozin:2016uqy}. These expressions for the master integrals are then included in the result for the renormalised correlator, and the entire expression is expanded using \texttt{HypExp} \cite{Huber:2005yg,Huber:2007dx} before taking the double discontinuity to calculate the spectral densities.

We add a few words of caution here.
The ``renormalised'' three-loop correlator is by no means renormalised in the sense of not having any $\frac{1}{\epsilon^n}$ poles.
In fact, its poles begin at $\frac{1}{\epsilon^3}$ even after renormalisation.
A simple argument shows that if the correlator has poles at most of order $\frac{1}{\epsilon^n}$ then the double discontinuity has poles of order $\frac{1}{\epsilon^{n-2}}$ or less.
This still leaves the door open for potentially disastrous simple $\frac{1}{\epsilon}$ poles in what we call the renormalised double discontinuity, i.e.\ the discontinuity of the renormalised correlator.
However, the $\frac{1}{\epsilon}$ poles in the double discontinuity stemming from the bare three-loop correlator and the renormalisation of the leading-order correlator cancel.
This is a non-trivial check of our calculation, since the former result from intrinsically non-factorisable expressions whereas the latter originate from the factorisable leading-order correlator.

\section{Details of the condensate calculation}\label{sec:condensates}
We also perform an analysis of the condensate contributions for all operators in the basis. 
For the SM operators, this has been done previously in the massless limit in Refs.~\cite{Cheng:1998ia,Baek:1998vk,Kirk:2017juj}, and also including strange quark mass effects in Ref.~\cite{King:2021jsq,King:Thesis22}.
This is tackled with the standard method of the background field technique~\cite{Novikov:1984ecy,Pascual:1984zb} and utilising the Fock-Schwinger gauge~\cite{Schwinger:1951nm,Fock:1937dy,Schwinger:1998lox} through which a fully-covariant expansion of the quark and gluon fields can be described such that condensates of increasing mass-dimension can be handled systematically. 

\begin{figure}[t]
    \centering
    \includegraphics[width=\textwidth]{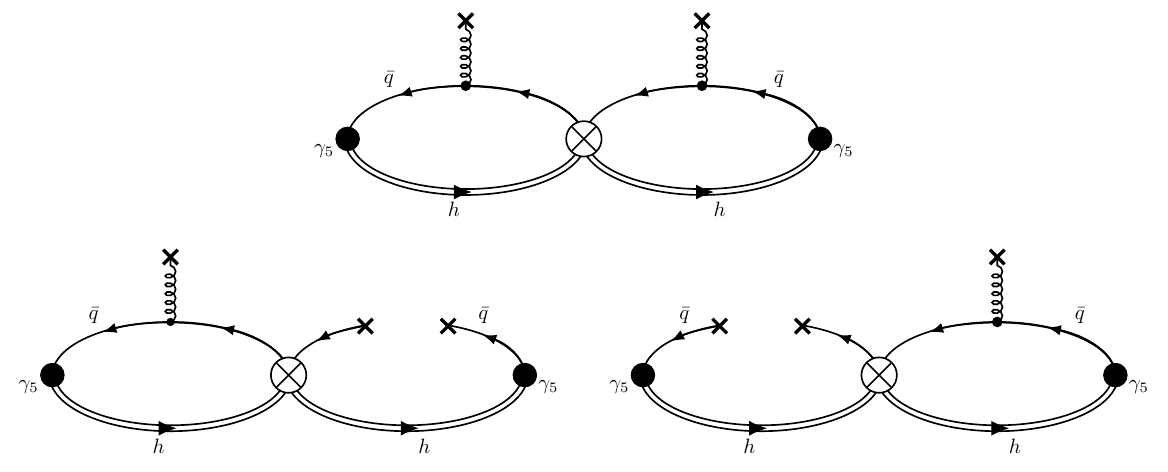}
    \caption{\label{fig:Cond_FD} Condensate contributions to the bag parameter beyond the VIA. The diagram on the top row is proportional to $\langle\frac{\alpha_s}{\pi}GG\rangle$ and those on the bottom row to $\langle g_s\bar{q}\sigma_{\mu\nu}G^{\mu\nu}q\rangle$.}
\end{figure}

Up to mass-dimension six and at leading order in $\alpha_s$, since the quark condensates are factorisable, the only condensates which enter the non-factorisable contribution are $\langle \frac{\alpha_s}{\pi}GG\rangle$ and $\langle g_s\bar{q}\sigma_{\mu\nu}G^{\mu\nu}q\rangle$; the diagrams corresponding to these are shown in \cref{fig:Cond_FD}.
To evaluate these contributions, one needs only the first correction to the light quark propagator in its expansion, corresponding to an emission of one gluon from the propagator; cf.\ \cref{eq:fermionpropexpandedfirstorder}. 
While one can also find similar condensate contributions from either expanding the quark fields or light quark propagator to higher orders, these will either add to the factorisable piece or only to the non-factorisable piece at higher orders in $\alpha_s$.

With the introduction of the corrected propagator, the evaluation of the $\langle \frac{\alpha_s}{\pi}GG\rangle$ diagram proceeds analogously to the procedure discussed above for the perturbative contributions, although this is technically simpler since it amounts to only a factorisable two-loop integral.
While the diagrams leading to the $\langle g_s\bar{q}\sigma_{\mu\nu}G^{\mu\nu}q\rangle$ contribution are rather similar and actually only contain a one-loop integral, they are conceptually slightly different. 
We therefore present an example calculation of this contribution in \cref{sec:condensateexample}.

\section{Results}\label{sec:results}
In this section we present the analytic and numerical results obtained for the perturbative and condensate contributions separately.

\subsection{Perturbative contribution}
Our results for the $r$-functions entering \cref{eq:DeltaBag} read
\begin{align}
    r_{\tilde{O}_1} \left(x, L_\nu\right) &= r_{\tilde{O}_2} \left(x, L_\nu\right) = r_{\tilde{O}_5} \left(x, L_\nu\right) = r_{\tilde{O}_6} \left(x, L_\nu\right) = 0 \,, \\
    r_{\tilde{O}_3} \left(x, L_\nu\right) &= r_{\tilde{T}_1} \left(x, L_\nu\right) = -7 + \frac{a_1}{8} + \frac{2\pi^2}{3} - \frac{3}{2} L_\nu - \frac{1}{4} \phi \left(x\right) \,, \label{eq:bagT1new} \\
    r_{\tilde{O}_4} \left(x, L_\nu\right) &= r_{\tilde{T}_2} \left(x, L_\nu\right) = -\frac{29}{4} + \frac{a_2}{8} + \frac{2\pi^2}{3} - \frac{3}{2} L_\nu - \frac{1}{4} \phi \left(x\right) \,, \label{eq:bagT2new} \\
    r_{\tilde{O}_7} \left(x, L_\nu\right) &= r_{\tilde{T}_3} \left(x, L_\nu\right) = 1 - \frac{a_3}{8} - \frac{2\pi^2}{3} \,, \label{eq:bagO7new} \\
    r_{\tilde{O}_8} \left(x, L_\nu\right) &= r_{\tilde{T}_4} \left(x, L_\nu\right) = \frac{15}{2} - \frac{2\pi^2}{3} + \frac{3}{2} L_\nu + \frac{1}{4} \phi \left(x\right) \,, \label{eq:bagO8new}
\end{align}
where
\begin{equation}
    L_\nu = \log\left(\frac{\mu^2}{4 \nu_1 \nu_2}\right) = \log\left(\frac{4\pi \mathrm{e}^{-\gamma_E} \tilde{\mu}^2}{4 \nu_1 \nu_2} \right) 
\end{equation}
and
\begin{equation}
    \phi \left(x\right) = \begin{cases} x^2 - 8x + 6 \log\left(x\right) \,, x \leq 1 \,, \\
    \frac{1}{x^2} - \frac{8}{x} - 6 \log\left(x\right) \,, x > 1 \end{cases} \,.
\end{equation}
For the SM operators $\tilde{O}_{1-4}$ these quantities have been computed in Ref.~\cite{Kirk:2017juj}.
We confirm the expression for $\tilde{O}_4 \equiv \tilde{T}_2$, but find a small difference for $\tilde{O}_3 \equiv \tilde{T}_1$ (the old constant of $-8$ is replaced by $-7$).
This updated expression, together with the entirely new contributions to the BSM operators $\tilde{O}_{5-8}$ are the main results of this work.

Unfortunately it was not possible to determine the exact cause for the difference between our new result, obtained by two completely independent calculations, and the one presented in Ref.~\cite{Kirk:2017juj}.
In this work, we present updated numerical bag parameters taking the shift in $r_{\tilde{T}_1} \left(x,L_\nu\right)$ into account.

We note that the function $\phi \left(x\right)$ is intrinsically non-factorisable and therefore cannot receive any contributions from the double discontinuity of the factorisable leading-order correlator through renormalisation.
Hence, it must be present and its coefficient must be finite already if one were to compute the double discontinuity of the \emph{bare} three-loop correlator.

\subsection{Condensate contribution}\label{subsec:condensates}
Since there is no gluon line like in the three-loop perturbative calculation, the operator insertion into the diagrams in \cref{fig:Cond_FD} must be a colour octet such that the colour trace is not zero, and thus the colour-singlet operators $\tilde{Q}_i$ have no condensate contribution.
Furthermore, the calculations for the colour-octet operators $\tilde{T}_{2,4}$ either reduce to scaleless integrals or have Dirac traces equal to zero.
Therefore the only operators for which the condensate diagrams are non-zero are $\tilde{T}_{1,3}$.
As one might expect from their Dirac structures, their contributions are equal.

In Ref.~\cite{Kirk:2017juj}, these terms were extracted from Ref.~\cite{Baek:1998vk}, where the interpolating current between the vacuum state and the valence-quark content of the meson has been taken as $\gamma_\mu \gamma_5$, which however interpolates $1^+$ states in addition to the desired $0^-$ states \cite{Cheng:1998ia}.
We use the pseudoscalar interpolating current instead \cite{Cheng:1998ia} and obtain vanishing condensate contributions to $\tilde{\epsilon}_{2,4}$.
This current has also been used in Ref.~\cite{King:2021jsq}, and we find full agreement of our analytical expressions for the condensate contributions.
With this interpolating current, we calculate the coefficients of the non-perturbative condensate matrix elements $\expval{\frac{\alpha_s}{\pi} GG}$ and $\expval{\bar{q} g_s \sigma_{\mu\nu} G^{\mu\nu} q}$, whose values have been determined in Refs.~\cite{Baek:1998vk,Kirk:2017juj} as
\begin{equation}\label{eq:condensatematrixelementsinput}
\expval{\frac{\alpha_s}{\pi} GG} = \SI{0.012 \pm 0.006}{\giga\eV}^4 \,, \qquad \expval{\bar{q} g_s \sigma_{\mu\nu} G^{\mu\nu} q} = -\SI{0.011 \pm 0.002}{\giga\eV}^5 \,.
\end{equation}

\begin{figure}[t]
    \centering
    \includegraphics[width=0.48\textwidth]{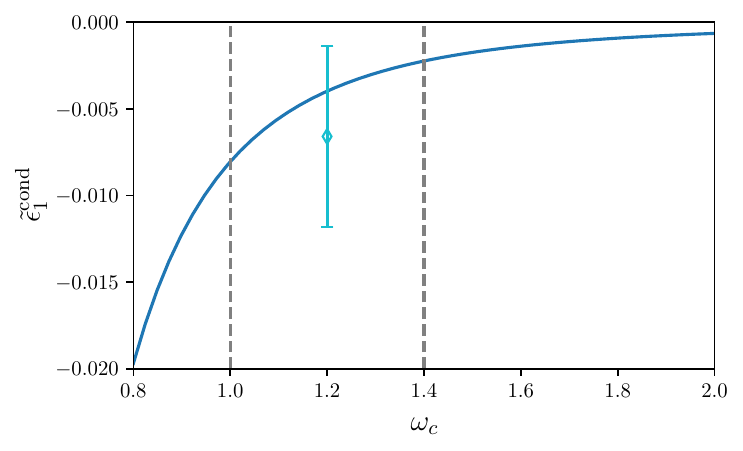}
    \includegraphics[width=0.48\textwidth]{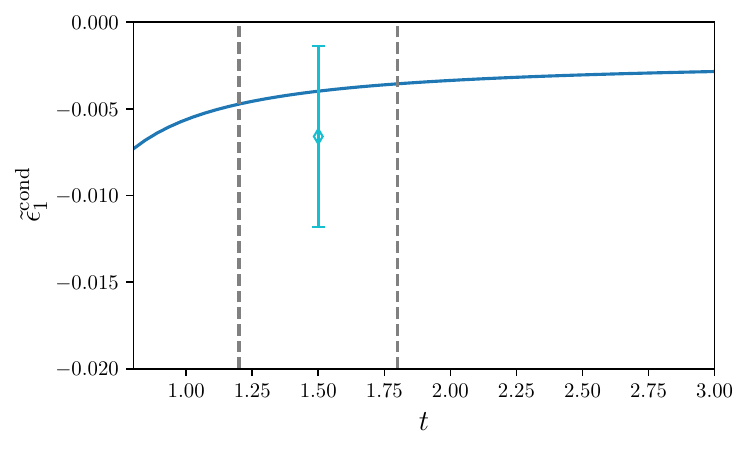}
    \caption{Variation of the condensate contribution to the bag parameter with the threshold energy $\omega_c$ (left) and the Borel parameter $t$ (right). The blue line shows the value as it is evolved for the single parameter, while the cyan data point indicates the final value after taking all variations into account. The dashed grey bands show the limits of the final ranges chosen.}
    \label{fig:condvars}
\end{figure}

In summary, the contributions to the non-factorisable part of the spectral density are therefore found to be
\begin{align}
    \label{eq:zerocond}
    \Delta\rho^{\text{cond}}_{\tilde{O}_i} &= 0, \qquad i=1,2,4,5,6,8, \\
    \Delta\rho^{\text{cond}}_{\tilde{T}_1} = \Delta\rho^{\text{cond}}_{\tilde{T}_3} &= -\frac{\langle\frac{\alpha_s}{\pi}GG\rangle}{64\pi^2} + \frac{N_c\langle g_s\bar{q}\sigma_{\mu\nu}G^{\mu\nu}q\rangle}{192\pi^2}\left[\delta(\nu_1)+\delta(\nu_2)\right].
\end{align}
Due to the presence of $\delta$-functions, we cannot reduce the condensate contribution to the bag parameters to something like \cref{eq:DeltaBag} as for the perturbative part and must instead use the traditional sum rule in \cref{eq:tradSR}.
We find the value of these contributions to be
\begin{equation}
\begin{alignedat}{3}
    \tilde{\epsilon}^{\text{cond}}_{1} &= -\tilde{\epsilon}^{\text{cond}}_{3} &&= - &&0.0067 \pm 0.0057 \,, \\
    \tilde{B}_i^{\text{cond}} &= \tilde{\epsilon}_{2}^{\text{cond}} = \tilde{\epsilon}_4^{\text{cond}} && = && 0.0000 \pm 0.0020 \,,
\end{alignedat}
\end{equation}
where we have varied the input parameters within their uncertainties to find the total spread in $\tilde{\epsilon}^{\text{cond}}$ and take a central value with symmetric errors. 
We have used $\mu_{\rho}=1.5\,$GeV and chose
\begin{align}
    \omega_c &= 1.2 \pm 0.2\,\text{GeV}, \\
    t_1 = t_2 \equiv t &= 1.5 \pm 0.3\,\text{GeV}.
\end{align}
To select these ranges, we vary the condensate contribution to the bag parameter in $\omega_c$ and $t$ and find the regions where these parameters are sufficiently small before the sum rule becomes unstable; these variations are shown in \cref{fig:condvars}.
An appropriate central value for $\omega_c$ is obtained by requiring that the sum rule appropriately overlaps with the ground state meson without contamination from the continuum contributions~\cite{Ball:1993xv,Lucha:2011zp,Mannel:2011iqd}.
Ultimately, there is a remaining degree of arbitrariness in the choice of $\omega_c$ which depends on the process at hand in the sense that it characterises the scale below which the OPE spectral density and the hadronic spectral density are identified.
We find that the bag parameter does not vary much with the Borel parameter $t$ in our chosen range, while it is more sensitive to $\omega_c$. 
Furthermore, in our assessment of the condensate uncertainties we have included an additional intrinsic condensate error of $\pm 0.002$ for all operators as an estimate of higher-dimensional condensates that have not been computed in this work.

\subsection{Bag parameters}
Combining the perturbative and condensate contributions presented in the previous subsections allows us to present numerical values for the bag parameters.
By \cref{eq:DeltaBag} the perturbative contribution to the bag parameter depends on the scale $\mu_\rho$ and on $\bar{\Lambda}$, which is typically a few hundred $\SI{}{\mega\eV}$.
The scale $\mu_\rho$ at which the bag parameters are evaluated should therefore be chosen close to $2 \bar{\Lambda}$ to avoid large logarithms.
At the same time, it should be a scale at which a perturbative definition of $\alpha_s \left(\mu_\rho\right)$ can still be justified, i.e.\ not too small either.
A reasonable compromise is the choice $\mu_{\rho} \sim \SI{1.5}{\giga\eV}$ and $\bar{\Lambda} \sim \SI{0.5}{\giga\eV}$ \cite{Kirk:2017juj}, which we vary in our uncertainty estimates by $\SI{100}{\mega\eV}$,
\begin{equation}\label{eq:barlambdarange}
    \bar{\Lambda} = \SI[separate-uncertainty = true]{0.5 \pm 0.1}{\giga\eV} \,.
\end{equation}
The value of $\alpha_s \left(\SI{1.5}{\giga\eV}\right)$ is determined from the strong coupling constant at the $Z$-boson mass, $\alpha_s \left( \SI{91.1880(20)}{\giga\eV}\right) = 0.1180(9)$ \cite{ParticleDataGroup:2024cfk}, which is evolved to the $b$-quark \MSbar{} mass $\overline{m}_b \left(\overline{m}_b \right) = \SI{4.183(7)}{\giga\eV}$ and decoupled with five-loop accuracy using \texttt{RunDec} \cite{Chetyrkin:2000yt,Schmidt:2012az,Herren:2017osy}.
The running of $\alpha_s$ from $\overline{m}_b \left(\overline{m}_b \right)$ to the scale at which the HQET bag parameter is then performed with two-loop accuracy, yielding a value of 
\begin{equation}
    \alpha_s \left(\SI{1.5}{\giga\eV}\right) = 0.3443 
\end{equation}
with all inputs being taken at their central values.
The uncertainty of $\alpha_s \left(M_Z\right)$ propagates into a $\sim 3 \, \%$ uncertainty of $\alpha_s \left(\mu_\rho\right)$, which is small compared to the other sources of uncertainty discussed in this section.
We will therefore neglect the $\alpha_s$ uncertainty in what follows.

In perturbative calculations, one typically tries to achieve scale independence through the computation of higher-order corrections, and the residual renormalisation-scale dependence as an estimate of neglected higher orders is assessed through variation of the scale $\mu$ in an interval $\left[\frac{1}{2} \mu_{\mathrm{central}}, 2 \mu_{\mathrm{central}}\right]$.
Unfortunately, the low overall energy scale of the HQET sum rule means that such a perturbative treatment with a lower value around $\mu_{\rho,\mathrm{min.}} \sim \SI{0.75}{\giga\eV}$ is not feasible.
Instead, we fix the scale at which the bag parameter is evaluated in HQET to be
\begin{equation}\label{eq:fixedmurho}
    \mu_\rho = \SI{1.5}{\giga\eV} 
\end{equation}
and estimate the uncertainty of neglected higher-order corrections by variation of $\mu_{\rho} \in \left[1,2\right]\SI{}{\giga\eV}$ and subsequently evolving back to the central scale $\mu_{\rho} = \SI{1.5}{\giga\eV}$ using the renormalisation-group (RG) equations.
Clearly, this is only a rough estimate, and in order to be conservative we inflate the associated uncertainty interval by a factor of $2$.

The anomalous dimension matrix in \cref{eq:HQETADMQQ} describes the RG evolution of the operators $\tilde{O}_i$ under renormalisation.
In the case at hand we need, however, the RG evolution of the bag parameters themselves, i.e.\ the anomalous dimension of the scale-dependent HQET decay constant has to be subtracted, such that~\cite{Kirk:2017juj}
\begin{equation}\label{eq:bagparamevolutionHQET}
    \vec{\tilde{B}} \left(\mu_1\right) = \left(\frac{\alpha_s \left(\mu_1\right)}{\alpha_s \left(\mu_0\right)} \right)^{\frac{\hat{\tilde{\gamma}}_{\tilde{B}}^{(0)}}{2 \beta_0}} \vec{\tilde{B}} \left(\mu_0\right) \,,
\end{equation}
with
\begin{equation}
    \hat{\tilde{\gamma}}_{\tilde{B}}^{(0)} = \left(\tilde{A}_{\tilde{O}}^D \right)^{-1} \hat{\tilde{\gamma}}_{\tilde{O}\tilde{O}} \tilde{A}_{\tilde{O}}^D -  2 \tilde{\gamma}_{\tilde{j}} \,. 
\end{equation}
Here, $\vec{B} \left(\mu\right)$ is the vector of all bag parameters and $\tilde{A}_{\tilde{O}}^D$ the diagonal matrix with the elements $\tilde{A}_i$ on its diagonal.
Due to the operator mixing under renormalisation all contributions to the bag parameters --- vacuum insertion approximation, perturbative contributions, and condensate terms --- have to be summed before assessing the $\mu_\rho$ scale uncertainty.

Finally, the use of \cref{eq:DeltaBag} rather than \cref{eq:tradSR} for the perturbative part of the bag parameter potentially introduces an additional uncertainty.
This issue has been discussed in Ref.~\cite{Kirk:2017juj}, where an additional intrinsic uncertainty of $\pm 0.02$ has been assigned to each bag parameter as an error estimate.
For our analysis we choose to keep the same intrinsic uncertainty.

We finally obtain the HQET bag parameters evaluated at $\mu_\rho = \SI{1.5}{\giga\eV}$, taking all evanescent scheme-dependent constants as $a_{1,2,3} = 0$,
\begin{equation}\label{eq:bagparametersHQETresults}
    \begin{alignedat}{9}
        &\tilde{B}_1 \left(\SI{1.5}{\giga\eV} \right) &&= && 1.0000 && \pm 0.0201 \\ 
        & && = && 1.0000 && \, _{-0.02}^{+0.02} \left(\mathrm{intr.}\right) && \,_{-0.0000}^{+0.0000} \left(\bar{\Lambda}\right) && \,_{-0.002}^{+0.002} \left(\mathrm{cond.}\right) && \,_{-0.0009}^{+0.0002} \left(\mu_\rho\right) \,, \\
        &\tilde{B}_2 \left(\SI{1.5}{\giga\eV} \right) &&= && 1.0000 && \pm 0.0201 \\
        & && = && 1.0000 && \, _{-0.02}^{+0.02} \left(\mathrm{intr.}\right) && \,_{-0.0000}^{+0.0000} \left(\bar{\Lambda}\right) && \,_{-0.002}^{+0.002} \left(\mathrm{cond.}\right) && \,_{-0.0011}^{+0.0000} \left(\mu_\rho\right) \,, \\
        &\tilde{\epsilon}_1 \left(\SI{1.5}{\giga\eV} \right) &&= - && 0.0053&&_{-0.0224}^{+0.0220} \\
        & && = - && 0.0053 && \, _{-0.02}^{+0.02} \left(\mathrm{intr.}\right) && \,_{-0.0082}^{+0.0067}\left(\bar{\Lambda}\right) && \,_{-0.0057}^{+0.0057} \left(\mathrm{cond.}\right) && \,_{-0.0018}^{+0.0029} \left(\mu_\rho\right) \,, \\
        &\tilde{\epsilon}_2 \left(\SI{1.5}{\giga\eV} \right) &&= - && 0.0017&&_{-0.0221}^{+0.0216} \\
        & && = - && 0.0017 && \, _{-0.02}^{+0.02} \left(\mathrm{intr.}\right) && \,_{-0.0082}^{+0.0067}\left(\bar{\Lambda}\right) && \,_{-0.002}^{+0.002} \left(\mathrm{cond.}\right) && \,_{-0.0041}^{+0.0042} \left(\mu_\rho\right) \,, \\
        &\tilde{B}_3 \left(\SI{1.5}{\giga\eV} \right) &&= && 1.0000 && \pm 0.0201 \\
        & && = && 1.0000 && \, _{-0.02}^{+0.02} \left(\mathrm{intr.}\right) && \,_{-0.0000}^{+0.0000} \left(\bar{\Lambda}\right) && \,_{-0.002}^{+0.002} \left(\mathrm{cond.}\right) && \,_{-0.0000}^{+0.0000} \left(\mu_\rho\right) \,, \\
        &\tilde{B}_4 \left(\SI{1.5}{\giga\eV} \right) &&= && 1.0000&&_{-0.0204}^{+0.0206} \\
        & && = && 1.0000 && \, _{-0.02}^{+0.02} \left(\mathrm{intr.}\right) && \,_{-0.0000}^{+0.0000} \left(\bar{\Lambda}\right) && \,_{-0.002}^{+0.002} \left(\mathrm{cond.}\right) && \,_{-0.0034}^{+0.0046} \left(\mu_\rho\right) \,, \\
        &\tilde{\epsilon}_3 \left(\SI{1.5}{\giga\eV} \right) &&= && 0.0747&&_{-0.0275}^{+0.0437} \\
        & && = && 0.0747 && \, _{-0.02}^{+0.02} \left(\mathrm{intr.}\right) && \,_{-0.0000}^{+0.0000} \left(\bar{\Lambda}\right) && \,_{-0.0057}^{+0.0057} \left(\mathrm{cond.}\right) && \,_{-0.0179}^{+0.0384} \left(\mu_\rho\right) \,, \\
        &\tilde{\epsilon}_4 \left(\SI{1.5}{\giga\eV} \right) &&= - && 0.0047&&_{-0.0217}^{+0.0212} \\
        & && = - && 0.0047 && \, _{-0.02}^{+0.02} \left(\mathrm{intr.}\right) && \,_{-0.0082}^{+0.0067} \left(\bar{\Lambda}\right) && \,_{-0.002}^{+0.002} \left(\mathrm{cond.}\right) && \,_{-0.0003}^{+0.0016} \left(\mu_\rho\right) \,.
    \end{alignedat}
\end{equation}
In the second line of each expression we have split the total uncertainty into its individual contributions from variation of $\bar{\Lambda}$ and $\mu_{\rho}$ as well as the additional condensate uncertainty and intrinsic sum rule uncertainty, respectively.

The SM bag parameters $\tilde{B}_{1,2}$ and $\tilde{\epsilon}_{1,2}$ have first been calculated in Ref.~\cite{Kirk:2017juj}, and subsequently strange-quark mass corrections and non-valence (eye) diagrams have been computed in Refs.~\cite{King:Thesis22,King:2021jsq}.
We confirm the numerical values for $\tilde{B}_{1,2}$ of Ref.~\cite{Kirk:2017juj}, while the values for $\tilde{\epsilon}_{1,2}$ differ.
The difference in the bag parameter $\tilde{\epsilon}_1$ is to some extent due to the larger negative value of the condensate contributions, but mainly due to the updated analytical expression in \cref{eq:bagT1new}.
Indeed, the perturbative contribution to $\tilde{\epsilon}_1$ now is positive at central input parameters, whereas it was negative before, as can be inferred by subtracting the condensate contributions from the total value of $\tilde{\epsilon}_1 \left(\SI{1.5}{\giga\eV}\right)$ both here and in Ref.~\cite{Kirk:2017juj}.
With the addition of the condensates, the total bag parameter $\tilde{\epsilon}_1 \left(\SI{1.5}{\giga\eV}\right)$ becomes negative again, but remains much closer to zero.
The difference in the bag parameter $\tilde{\epsilon}_2$, on the other hand, is due to a different treatment of condensate effects; see the discussion in \cref{subsec:condensates}.
Despite the differences compared to Ref.~\cite{Kirk:2017juj} we emphasise that our new central values for the bag parameters $\tilde{\epsilon}_{1,2}$ are within the uncertainties quoted in that publication.

Finally, we emphasise that the total uncertainties of all bag parameters with the exception of $\tilde{\epsilon}_3$ are dominated by the intrinsic sum rules uncertainty of $0.02$.
In particular, for the colour-octet bag parameters except $\tilde{\epsilon}_3$, the uncertainties are much larger than the magnitude of the bag parameters themselves.
This is in contrast to the results of Ref.~\cite{Kirk:2017juj}, where $\tilde{\epsilon}_1 \left(\SI{1.5}{\giga\eV}\right) = -0.016_{-0.022}^{+0.021}$.
It is possible that the intrinsic error of $\pm 0.02$ is a severe overestimation of the sum rule uncertainties, yet at the same time we cannot fully exclude slightly larger uncertainties for $\tilde{\epsilon}_3$.
We therefore do not attempt to argue in favour of a smaller intrinsic uncertainty here.

The bag parameters presented here can be used in order to determine the SM lifetime ratio $\tau\left(B^+\right) / \tau\left(B_d\right)$ using the Wilson coefficients computed in Refs.~\cite{Ciuchini:2001vx,Beneke:2002rj,Franco:2002fc}.
In anticipation of a comprehensive numerical study of $B$-meson lifetimes \cite{Egner:2024lay} including  the recently computed NNLO  corrections to $\Gamma_3$ \cite{Egner:2024azu}, we refrain from updating the lifetime ratios here and refer to the upcoming publication, but we expect a slight decrease of $\tau\left(B^+\right) / \tau\left(B_d\right)$ towards the experimental value.

\subsection{HQET--QCD matching for the Standard Model operators}
We also convert the HQET bag parameters to their QCD equivalents for the SM operators defined in \cref{eq:QCDphysops,eq:IBopsQCD}, in order to facilitate comparison with existing results.

The QCD bag parameters are evaluated at a scale more natural to $B$-meson decays, which we choose to be $\overline{m}_b \left(\overline{m}_b\right)$.
In order to convert the bag parameters from HQET to QCD, a matching step has to be performed at some scale $\mu_m$ typically somewhere between $\mu_\rho = \SI{1.5}{\giga\eV}$ and $\mu_Q = \overline{m}_b \left(\overline{m}_b\right)$ in order to avoid large logarithms on either side of the matching calculation.
The matching matrix $C_{O_i \tilde{O}_j} \left(\mu\right)$ between $\left\{\tilde{Q}_1, \tilde{Q}_2, \tilde{T}_1, \tilde{T}_2 \right\}$ and $\left\{Q_1,Q_2,T_1,T_2\right\}$ has been computed including $\mathcal{O}\left(\alpha_s\right)$ corrections in Ref.~\cite{Kirk:2017juj}.
Thus, the bag parameters in QCD are determined by evolving the HQET bag parameters from $\SI{1.5}{\giga\eV}$ to $\mu_m$ using \cref{eq:bagparamevolutionHQET}, performing the matching at the scale $\mu_m$ by use of the relation\footnote{Here $B$ denotes the bag parameter of any colour-singlet or colour-octet operator.}\cite{Kirk:2017juj}
\begin{equation}
    B_{O_i} \left(\mu_m\right) = \sum_{j} \frac{\tilde{A}_j}{A_i} \frac{C_{O_i \tilde{O}_j} \left(\mu_m\right)}{\left(C\left(\mu_m\right)\right)^2} \tilde{B}_j \left(\mu_m\right) + \mathcal{O} \left(\frac{1}{m_b}\right) \,,
\end{equation}
with
\begin{equation}
    C \left(\mu_m\right) = 1 - 2 C_F \frac{\alpha_s \left(\mu_m\right)}{4\pi} + \mathcal{O}\left(\alpha_s^2\right) \,,
\end{equation}
and subsequently running the bag parameters from $\mu_m$ to $\overline{m}_b \left(\overline{m}_b\right)$ using the QCD version of \cref{eq:bagparamevolutionHQET} which is analogous except for the absence of the $-2 \tilde{\gamma}_{\tilde{j}}$ term in the anomalous dimension matrix, since the leptonic decay constants are scale independent in QCD.
Since the HQET NLO anomalous dimension matrix of the operators, $\tilde{\gamma}_{\tilde{O}\tilde{O}}^{(1)}$, is currently not known, we cannot fully perform the matching at NLO.
Instead, we use the LO anomalous dimension matrices on both the HQET and QCD sides for the RG evolution, while we perform the matching itself at $\mathcal{O}\left(\alpha_s\right)$.
Finally, we relate the on-shell mass $m_b^{\mathrm{OS}}$ in \cref{eq:afactorsqcd} to the meson mass using the relation
\begin{equation}\label{eq:mbOSLambdaRelation}
    m_b^{\mathrm{OS}} = M_{B_d} - \bar{\Lambda} + \mathcal{O}\left(\frac{1}{m_b}\right) \,,
\end{equation}
and expand the factor $A_2$ strictly as $A_2 = 1 + \mathcal{O}\left(1/m_b\right)$.
The matching scale $\mu_m$ is varied between $\left[3,6\right]\SI{}{\giga\eV}$ and we allow the scheme constants $a_1, a_2$ of the evanescent HQET operators to float between $\left[-10,10\right]$ in order to account for the neglected NLO corrections to the RG evolution and additional higher-order perturbative corrections to the relation between $B_i\left(\overline{m}_b \left(\overline{m}_b\right)\right)$ and $\tilde{B}_i \left(\SI{1.5}{\giga\eV}\right)$.
The ``QCD choice'' corresponds to $a_1 = a_2 = -8$, but $a_3 = -16$; however, for the matching of the SM operators $a_3$ is irrelevant.
Finally, we perform a ``scheme conversion'' such that the matrix elements $\overline{\expval{Q_2}}, \overline{\expval{T_2}}$ are expressed as
\begin{align}
    \overline{\expval{Q_2}}_{B^+} \left(\mu\right) &= \overline{A}_2 \left(\mu\right) f_{B^+}^2 M_{B^+}^2 \overline{B}_2 \left(\mu\right) \,, \\
    \overline{\expval{T_2}}_{B^+} \left(\mu\right) &= \overline{A}_2 \left(\mu\right) f_{B^+}^2 M_{B^+}^2 \overline{\epsilon}_2 \left(\mu\right) \,,
\end{align}
with
\begin{equation}
    \overline{A}_2 \left(\mu\right) = \frac{M_{B^+}^2}{\left(\overline{m}_b \left(\mu\right)\right)^2} \,.
\end{equation}
We find the bag parameters in QCD to be
\begin{equation}\label{eq:bagparametersQCDresults}
    \begin{alignedat}{8}
        & \overline{B}_1 \left(\mbmb\right) &&= && 1.013_{-0.059}^{+0.066} &&= && 1.013 &&\, _{-0.028}^{+0.028} (\text{HQET S.R.}) && \, _{-0.052}^{+0.059} (\text{match.}) \,, \\
        & \overline{B}_2 \left(\mbmb\right) &&= && 1.004_{-0.081}^{+0.085} &&= && 1.004 &&\, _{-0.024}^{+0.024} (\text{HQET S.R.}) && \, _{-0.077}^{+0.082} (\text{match.}) \,, \\
        & \overline{\epsilon}_1 \left(\mbmb\right) &&= - && 0.098_{-0.032}^{+0.029} &&= - && 0.098  &&\, _{-0.024}^{+0.024} (\text{HQET S.R.}) && \, _{-0.021}^{+0.016} (\text{match.}) \,, \\
        & \overline{\epsilon}_2 \left(\mbmb\right) &&= - && 0.037_{-0.020}^{+0.019} &&= - && 0.037 &&\, _{-0.016}^{+0.016} (\text{HQET S.R.}) && \, _{-0.013}^{+0.010} (\text{match.}) \,.
    \end{alignedat}
\end{equation}
We emphasise that despite the significant changes of the HQET bag parameters $\tilde{\epsilon}_1, \tilde{\epsilon}_2$, the QCD bag parameters at the scale $\mbmb$ are very close to those determined in Ref.~\cite{Kirk:2017juj} and well within the uncertainties quoted in Ref.~\cite{Kirk:2017juj}.
However the $\mathcal{O}\left(10 \,\%\right)$ deviations of $\overline{\epsilon}_{1,2}$ with respect to the results of Ref.~\cite{Kirk:2017juj} deserve an explanation.
We find that the central values of $\overline{B}_1 \left(\mbmb\right)$ and $\overline{B}_2 \left(\mbmb\right)$ differ by less than $\sim 1.6\,\%$ from those of Ref.~\cite{Kirk:2017juj}, with slightly larger uncertainties.
This difference is mostly due to slightly different input values translating into differences of about this size when performing the matching and RG evolution.
The parameters $\bar{\epsilon}_{1,2} \left(\mbmb\right)$ on the other hand are dominated not by the values $\tilde{\epsilon}_{1,2} \left(\SI{1.5}{\giga\eV}\right)$, but by $\tilde{B}_{1,2} \left(\SI{1.5}{\giga\eV}\right)$ through operator mixing under RG evolution and the HQET-QCD matching.
A tiny difference in the matching-evolution matrix can therefore lead to a sizeable relative deviation of $\bar{\epsilon}_{1,2} \left(\mbmb\right)$.
Finally, for $\bar{\epsilon}_2$, part of the difference also stems from the updated treatment of the condensates; see \cref{sec:condensates}.

We are careful to keep our expressions through the matching procedure symbolic until the very end where we truncate at linear order in $\alpha_s$ before evaluating. 
If one evaluates the components of the matching equation independently before combining, higher-order $\alpha_s$ contributions will enter the final values, however as these are not the complete NNLO (or beyond) picture, we decide to truncate these terms.
While this does not lead to extreme numerical effects in the matching for $B$ mesons discussed here, it is significant in the matching for $D$ mesons discussed below.

We also perform the HQET-QCD matching for the case of $D$ mesons which we consider at the scale $\mu=3\,$GeV.
We find
\begin{equation}\label{eq:bagparametersQCDresultscharm}
    \begin{alignedat}{8}
        & \overline{B}_1 \left(3\,\text{GeV}\right) &&= && 0.875_{-0.044}^{+0.070} &&= && 0.875 &&\, _{-0.027}^{+0.028} (\text{HQET S.R.}) && \, _{-0.035}^{+0.064} (\text{match.}) \,, \\
        & \overline{B}_2 \left(3\,\text{GeV}\right) &&= && 0.862_{-0.078}^{+0.138} &&= && 0.862 &&\, _{-0.018}^{+0.018} (\text{HQET S.R.}) && \, _{-0.076}^{+0.137} (\text{match.}) \,, \\
        & \overline{\epsilon}_1 \left(3\,\text{GeV}\right) &&= - && 0.122_{-0.042}^{+0.033} &&= - && 0.122  &&\, _{-0.027}^{+0.027} (\text{HQET S.R.}) && \, _{-0.033}^{+0.019} (\text{match.}) \,, \\
        & \overline{\epsilon}_2 \left(3\,\text{GeV}\right) &&= && 0.0002_{-0.0197}^{+0.0148} &&= && 0.0002 &&\, _{-0.0094}^{+0.0092} (\text{HQET S.R.}) && \, _{-0.0173}^{+0.0115} (\text{match.}) \,.
    \end{alignedat}
\end{equation}
We have evaluated the matching uncertainty by varying the HQET-QCD matching scale in the interval $\mu_m \in \left[2,4\right]\SI{}{\giga\eV}$ and allowing the scheme constants $a_{1,2}$ within $\left[-10,10\right]$. 
These values agree with the previous results of Ref.~\cite{Kirk:2017juj} within uncertainties, but the coefficient $\overline{B}_2$ deserves attention and will be discussed below.
Similarly to the discussion above for $B$ mesons, we find ${\cal O}(10\,\%)$ deviations with respect to the previous results, which can largely be accounted for through the relatively small changes in the HQET bag parameters being amplified in the matching to QCD yielding larger differences.
Furthermore, we find that the HQET-QCD matching for charm quarks is much more sensitive to the exact perturbative treatment in the numerical assessment discussed above. 
Due to the larger size of $\alpha_s(\SI{3}{\giga\eV})$ and more importantly, terms proportional to $\log\left(\mu^2/\mcmc^2\right)$ appearing, the variations of the final values, if one is not consistent with the order of truncation, can be significant. 
As discussed above for the $B$ mesons, we again keep our expressions symbolic in $\alpha_s$ until the conversion to the \MSbar{} scheme for the prefactor $A_2$ has been performed.
With this choice no partial $\mathcal{O}\left(\alpha_s^2\right)$ terms are included in $\overline{B}_i \left(\SI{3}{\giga\eV}\right)$, which is the main reason why our central value for $\overline{B}_2 \left(\SI{3}{\giga\eV}\right)$ is larger than the one quoted in Ref.~\cite{Kirk:2017juj} by almost $17\,\%$.
If we numerically evaluate the bag parameters before performing the conversion of $A_2$, i.e.\ keep partial $\mathcal{O}\left(\alpha_s^2\right)$ results, we find a value very close to the one published in Ref.~\cite{Kirk:2017juj}, and hence we expect this difference in perturbative treatment to be the source of discrepancy.
It is interesting to note that our updated central value is at the very edge of the uncertainty estimate for $\overline{B}_2 \left(\SI{3}{\giga\eV}\right)$ presented in Ref.~\cite{Kirk:2017juj}, while our lower uncertainty does not reach all the way down to the old central value.
This clearly shows that the naive estimate for the matching uncertainty obtained via scale and scheme constant variation appears to be at best a lower bound for the total uncertainty.

In both the $B$-meson and $D$-meson cases improvement could possibly be achieved by a calculation of the NLO HQET anomalous dimension $\hat{\tilde{\gamma}}_{\tilde{O}\tilde{O}}^{(1)}$, which would allow for a fully consistent matching and RG evolution, as well as higher-order perturbative corrections to the matching matrix itself.
Still, due to the bad convergence behaviour of the perturbative $m_c^{\mathrm{OS}} / \overline{m}_c \left(\overline{m}_c\right)$ relation a  precise determination of the bag parameters $\overline{B}_i$  for $D$ mesons within the HQET sum rule framework will be very difficult.

\section{Conclusion}\label{sec:conclusion}
We have computed the hadronic matrix elements of the BSM $\Delta B = 0$ effective HQET dimension-six four-quark operators within the framework of HQET sum rules.
More precisely, we have considered the isospin-breaking operator combinations $\tilde{O}_i^u - \tilde{O}_i^d$ that are relevant for the description of lifetime ratios.
The technically most challenging part of the calculation involved the evaluation of three-loop HQET diagrams, but the condensate contributions play an equally important role numerically.
As an intermediate step we have recomputed the previously known bag parameters for the SM subset of operators $\left\{\tilde{Q}_{1}, \tilde{Q}_2, \tilde{T}_1, \tilde{T}_2\right\}$ and found a difference in the perturbative part compared to the original publication \cite{Kirk:2017juj}, leading to an updated SM prediction for the lifetime ratio of charged to neutral $B$ mesons \cite{Egner:2024lay}.
In general, we find all bag parameters to be very close to the vacuum insertion approximation.

Our results for the BSM operators serve as an input for further studies within specific scenarios of new physics (NP).
The $\Delta B = 0$ short-distance coefficients have been computed in Ref.~\cite{Lenz:2022pgw} in terms of the Wilson coefficients $C_i$ of the most general HQET effective $\abs{\Delta B} = 1$ Hamiltonian relevant for meson lifetimes.
Within a definite ultraviolet scenario that specifies the coefficients $C_i$, these $\Delta B = 0$ coefficients can be combined with the hadronic matrix elements presented in this work in order to obtain quantitative predictions for the lifetime ratio.

While older lattice results are available in Refs.~\cite{DiPierro:1998ty,DiPierro:1999tb,Becirevic:2001fy}, we do not directly compare to these as they were early studies performed on quenched ($n_f=0$) lattices where the error stemming from this quenched approximation cannot be quantified. 
More recently, Ref.~\cite{Black:2024iwb} presents preliminary values for the bag parameters $\bar{B}_1(\SI{3}{\giga\eV})$ and $\bar{\epsilon}_1(\SI{3}{\giga\eV})$ for the $D_s$ meson system while the matching to $\MSbar{}$ assumes lifetime-difference operators. 
These values differ by $\sim2-3\sigma$ from those quoted in \cref{eq:bagparametersQCDresultscharm}, however one must caveat this with the presence of the strange spectator quark in the lattice calculation which may cause such differences. 
Future updates on the lattice studies in Refs.~\cite{Lin:2022fun,Black:2024iwb} are expected to result in values for both $B$ and $D$ meson systems with both light and strange spectator quarks where more direct comparisons to our results here can then be made.

\section*{Acknowledgements}
The authors would like to thank K.\ Brune, M.\ Egner, T.\ Huber, A.\ Khodjamirian, F.\ Lange, M.L.\ Piscopo, and A.\ Rusov for helpful discussions and M.\ Kirk for help in the attempt to understand the difference between our result an the ones in Ref.~\cite{Kirk:2017juj}.
We would also like to thank M.\ Steinhauser for providing helpful information about the program \texttt{RunDec}.
This project was supported by the Deutsche Forschungsgemeinschaft (DFG, German Research Foundation) under grant 396021762-TRR 257
and the BMBF project ``Theoretische Methoden für LHCb und Belle II''  (Förderkennzeichen 05H21PSCLA/ErUM-FSP T04).
M.B.\ was additionally funded in part by UK STFC grant ST/X000494/1.

\appendix

\section{Evanescent operators and anomalous dimension matrix in QCD}\label{sec:evanQCD}
Here we list the operators needed to complete the set of physical operators in QCD at one loop.
They are given as \cite{Kirk:2017juj,Aebischer:2022aze}
\begin{equation}\label{eq:QCDevanescentoperators}
    \begin{aligned}
       E_{1}^q &\equiv \bar{b} \gamma_{\mu\nu\rho} \left(1 - \gamma_5\right) q \, \bar{q} \gamma^{\rho\nu\mu} \left(1 - \gamma_5\right) b \quad - \left(4 - 8 \epsilon\right) O_{1}^q \,, \\
       E_{2}^q &\equiv \bar{b} \gamma_{\mu\nu} \left(1 - \gamma_5\right) q \, \bar{q} \gamma^{\nu\mu} \left(1 + \gamma_5\right) b \quad - \left(4 - 8 \epsilon\right) O_{2}^q \,, \\
       E_{3}^q &\equiv \bar{b} \gamma_{\mu\nu\rho} \left(1 - \gamma_5\right) T^a q \, \bar{q} \gamma^{\rho\nu\mu} \left(1 - \gamma_5\right) T^a b \quad - \left(4 - 8 \epsilon\right) O_{3}^q \,, \\
       E_{4}^q &\equiv \bar{b} \gamma_{\mu\nu} \left(1 - \gamma_5\right) T^a q \, \bar{q} \gamma^{\nu\mu} \left(1 + \gamma_5\right) T^a b \quad - \left(4 - 8 \epsilon\right) O_{4}^q \,, \\
       E_{5}^q &\equiv \bar{b} \gamma_{\mu\nu\rho} \left(1 - \gamma_5\right) q \, \bar{q} \gamma^{\rho\nu\mu} \left(1 + \gamma_5\right) b \quad - \left(16 - 16\epsilon \right) O_5^q \,, \\
       E_{6}^q &\equiv \bar{b} \gamma_{\mu\nu\rho} \left(1 - \gamma_5\right) T^a q \, \bar{q} \gamma^{\rho\nu\mu} \left(1 + \gamma_5\right) T^a b \quad - \left(16 - 16\epsilon \right) O_7^q \,, \\
      E_{7}^q &\equiv \bar{b} \sigma_{\mu\nu} \gamma_{\rho\sigma} \left(1 - \gamma_5\right) q \, \bar{q} \gamma^{\sigma\rho} \sigma^{\mu\nu} \left(1 - \gamma_5\right) b \\
      & \qquad - \left(48 - 80 \epsilon \right) O_6^q \quad - \left(12 - 14 \epsilon\right) O_{9}^q \,, \\
      E_{8}^q &\equiv \bar{b} \sigma_{\mu\nu} \gamma_{\rho\sigma} \left(1 - \gamma_5\right) T^a q \, \bar{q} \gamma^{\sigma\rho} \sigma^{\mu\nu} \left(1 - \gamma_5\right) T^a b \\
      & \qquad - \left(48 - 80 \epsilon \right) O_8^q \quad - \left(12 - 14 \epsilon\right) O_{10}^q \,,
    \end{aligned}
\end{equation}
plus the corresponding primed counterparts.

For the QCD operators, the anomalous dimension matrices of the physical operators have been computed in Refs.~\cite{Ciuchini:1997bw,Buras:2000if}. In our basis with operators $O_{i=1,\ldots,10}^q$ and their respective evanescent operators $E_{i=1,\ldots,8}^q$, they are given by\footnote{We recomputed them directly with vanishing external momenta and universal mass $m$ for all quark propagator denominators and checked agreement with Refs.~\cite{Ciuchini:1997bw,Buras:2000if}. The gluon propagator does not need IR regularisation in this case.}
\begin{equation}\scriptsize
    \hat{\gamma}_{OO}^{(0)} = \begin{pmatrix} 
    0 & 0 & 12 & 0 & 0 & 0 & 0 & 0 & 0 & 0 \\
    0 & \frac{6}{N_c} - 6 N_c & 0 & 0 & 0 & 0 & 0 & 0 & 0 & 0 \\
    3 - \frac{3}{N_c^2} & 0 & -\frac{12}{N_c} & 0 & 0 & 0 & 0 & 0 & 0 & 0\\
    0 & 0 & 0 & \frac{6}{N_c} & 0 & 0 & 0 & 0 & 0 & 0\\
    0 & 0 & 0 & 0 & 0 & 0 & -12 & 0 & 0 & 0 \\
    0 & 0 & 0 & 0 & 0 & \frac{6}{N_c} - 6 N_c & 0 & 0 & 0 & -2 \\
    0 & 0 & 0 & 0 & -3 + \frac{3}{N_c^2} & 0 & \frac{12}{N_c} - 6 N_c & 0 & 0 & 0 \\
    0 & 0 & 0 & 0 & 0 & 0 & 0 & \frac{6}{N_c} & -\frac{1}{2} \left(1 - \frac{1}{N_c^2}\right) & \frac{2}{N_c} - \frac{N_c}{2} \\
    0 & 0 & 0 & 0 & 0 & 0 & 0 & -96 & -\frac{2}{N_c} + 2 N_c & 0 \\
    0 & 0 & 0 & 0 & 0 & 24\left(-1+\frac{1}{N_c^2}\right) & 0 & \frac{96}{N_c} - 24 N_c & 0 & -\frac{2}{N_c} - 4 N_c
    \end{pmatrix}
\end{equation}
and
\begin{equation}\label{eq:QCDADMQE}\scriptsize
    \hat{\gamma}_{OE}^{(0)} = \begin{pmatrix}
    0 & 0 & -2 & 0 & 0 & 0 & 0 & 0\\
    0 & 0 & 0 & -2 & 0 & 0 & 0 & 0\\
    -\frac{1}{2} + \frac{1}{2N_c^2} & 0 & \frac{2}{N_c} - \frac{N_c}{2} & 0 & 0 & 0 & 0 & 0\\
    0 & -\frac{1}{2} + \frac{1}{2N_c^2} & 0 & \frac{2}{N_c} - \frac{N_c}{2} & 0 & 0 & 0 & 0\\
    0 & 0 & 0 & 0 & 0 & -2 & 0 & 0\\
    0 & 0 & 0 & 0 & 0 & 0 & 0 & 0\\
    0 & 0 & 0 & 0 & -\frac{1}{2} + \frac{1}{2N_c^2} & \frac{2}{N_c} - \frac{N_c}{2} & 0 & 0 \\
    0 & 0 & 0 & 0 & 0 & 0 & 0 & 0 \\
    0 & 0 & 0 & 0 & 0 & 0 & 0 & -2 \\
    0 & 0 & 0 & 0 & 0 & 0 & -\frac{1}{2} + \frac{1}{2N_c^2} & \frac{2}{N_c} - \frac{N_c}{2}
    \end{pmatrix}
\end{equation}
for the evanescent operators.
Our result for the upper left (SM) subblock of \cref{eq:QCDADMQE} agrees with Ref.~\cite{Kirk:2017juj}.
The anomalous dimension matrices for the primed operators $O_{i}^{q\prime}, E_{i}^{q\prime}$ are identical to $\gamma_{QQ}^{(0)}, \gamma_{QE}^{(0)}$ since QCD preserves chirality.

\section{Note on the Double Discontinuity}\label{sec:DoubleDisc}

Analyticity of the S-matrix requires the correlator $K(\omega_1,\omega_2)$ to be a complex analytic function of its variables, and we calculate $K$ perturbatively in the unphysical regime $\omega_1,\,\omega_2<0$. 
In this region the correlator is analytic, whereas it has a cut for $\omega_1 > 0$ and a cut for $\omega_2>0$.
The correlator is related to the physical regime via two implementations of Cauchy's integral formula, with Pac-Man-like contour integrals. 
We can ignore the integration over the circular parts of the contour integral, as the Borel transformation later in the calculation will remove any remaining artefacts of these. 
In the end, it is left to take the two discontinuities of the correlator over the two branch cuts.

We define the first discontinuity as
\begin{equation}
    \begin{aligned}
    \rho_1(\omega_1,\omega_2) = \frac{1}{2\pi\mathrm{i}} \left[K(\omega_1+\mathrm{i}\alpha,\omega_2)-K(\omega_1-\mathrm{i}\alpha,\omega_2) \right]
    \end{aligned}
\end{equation}
where we hold $\omega_2<0$ and take $\omega_1>0$ to be on the branch cut, and $\alpha>0$ is to be taken to go to zero. 

The branch cut can be completely described by functions of the form $\left\{ (-\omega)^{2-3\epsilon},\,\right.$ $\left. \log\left(-\omega\right),\, \log\left(x\right),\, \text{Li}_2\left(1-x\right),\, \text{Li}_3\left(1-x\right),\, \text{Li}_2\left(1-1/x\right)\right\}$, where $\omega$ can be either $\omega_1$ or $\omega_2$ and $x=\omega_2/\omega_1$. After adding and subtracting $\mathrm{i}\alpha$, we implement the following replacements  to take the discontinuity:
\begin{equation}
    \begin{aligned}
    z^a &\rightarrow
    \begin{cases}
    e^{\mathrm{i} \pi a} (-z)^a & \text{if } \Re{z}<0 \text{ and } \Im{z}>0,\\
    e^{-\mathrm{i} \pi a} (-z)^a & \text{if } \Re{z}<0 \text{ and } \Im{z}<0,\\
    z^a & \text{else,}
    \end{cases}
    \\
    \log\left(z\right) &\rightarrow 
    \begin{cases}
    \log\left(-z\right) + \mathrm{i}\pi & \text{if } \Re{z}<0 \text{ and } \Im{z}>0,\\
    \log\left(-z\right) - \mathrm{i}\pi & \text{if } \Re{z}<0 \text{ and } \Im{z}<0,\\
    \log\left(z\right) & \text{else,}
    \end{cases}
    \\
    \text{Li}_n\left(z\right) &\rightarrow 
    \begin{cases}
    \Re{\text{Li}_n\left(z\right)} + \frac{\mathrm{i}\pi}{\Gamma\left(n\right)}\log^{n-1}\left(z\right) & \text{if } \Re{z}<0 \text{ and } \Im{z}>0,\\
     \Re{\text{Li}_n\left(z\right)} - \frac{\mathrm{i}\pi}{\Gamma\left(n\right)}\log^{n-1}\left(z\right) & \text{if } \Re{z}<0 \text{ and } \Im{z}<0,\\
    \text{Li}_n\left(z\right) & \text{else}\,.
    \end{cases}
    \end{aligned}
\end{equation}
After the replacements are made, we take $\alpha\to0$. 
There is an important subtlety here: as we need to take the discontinuity again, these replacement functions must be analytic on the entire complex plane. 
Thus, as the term $\log\left(x\right)\text{Li}_2\left(1-x\right)$ appears in the correlator, we require a formula for the analytic continuation of the real part of the polylogarithm function along the branch cut. For this we use
\begin{equation}
    \begin{aligned}
    \Re{\text{Li}_2\left(z\right)}= \pi^2/6-\log\left(z-1\right)\log\left(z\right)-\text{Li}_2\left(1-z\right) \text{, for } z>1 \,,
    \end{aligned}
\end{equation}
which is easily derived from one of Euler's identities for polylogarithms. As the right-hand side of this equation is complex analytic over the entire complex plane (except for the branch cut for $z<1$), and agrees with the left-hand side for $z>1$, we have found the analytic continuation.

The second discontinuity is then defined by
\begin{equation}
    \begin{aligned}
   \rho_{\tilde{O}_i} = \rho_2\left(\omega_1,\omega_2\right) = \frac{1}{2\pi\mathrm{i}} \left[\rho_1(\omega_1,\omega_2+\mathrm{i}\alpha)-\rho_1(\omega_1,\omega_2-\mathrm{i}\alpha) \right]
    \end{aligned}
\end{equation}
where now we hold $\omega_1>0$ and assume $\omega_2>0$ is on the branch cut. The same replacements are then used as for the first discontinuity.

\section{Master Integrals}\label{sec:masterintegrals}
Using IBP reduction all Feynman integrals occurring in the three-loop perturbative calculation could be reduced to a set of $11$ master integrals, 
\begin{equation}
    \begin{aligned}
        &I_a \left(0, 1, 1, 0, 0, 1, 1; 0, 0; \omega_1, \omega_2 \right) \,, \\
        &I_a \left(1, 0, 0, 1, 0, 1, 1; 0, 0; \omega_1, \omega_2 \right) \,, \\ 
        &I_a \left(1, 1, 0, 0, 1, 1, 1; 0, 0; \omega_1, \omega_2 \right) \,, \\
        &I_a \left(1, 1, 0, 0, 2, 1, 1; 0, 0; \omega_1, \omega_2 \right) \,, \\
        &I_a \left(1, 1, 0, 1, 0, 1, 1; 0, 0; \omega_1, \omega_2 \right) \,, \\
        &I_a \left(1, 1, 0, 1, 1, 1, 0; 0, 0; \omega_1, \omega_2 \right) \,, \\
        &I_a \left(1, 1, 1, 0, 0, 1, 1; 0, 0; \omega_1, \omega_2 \right) \,, \\
        &I_a \left(1, 1, 1, 0, 1, 0, 1; 0, 0; \omega_1, \omega_2 \right) \,, \\
        &I_a \left(1, 1, 1, 1, 0, 1, 1; 0, 0; \omega_1, \omega_2 \right) \,, \\
        &I_b \left(1, 1, 1, 1, 1, 0, 1; 0, 0; \omega_1, \omega_2 \right) \,, \\
        &I_{\tilde{b}} \left(1, 1, 1, 1, 1, 0, 1; 0, 0; \omega_1, \omega_2 \right)\,,
    \end{aligned}
\end{equation}
as defined in Refs.~\cite{Grozin:2008nu,Grozin:2016uqy} with $I_{\tilde{b}} = I_b \left(\omega_1 \leftrightarrow \omega_2\right)$, where they have been computed and expanded in $\epsilon$.
{Note that the $\epsilon^4$ term in the expansion of the master integral $M_3$ in Ref.~\cite{Grozin:2016uqy} (first equation of appendix A) contains a typo: the term $+288 L^2\left(x\right)$ should read $+288 x L^2 \left(x\right)$.}

\section{Anomalous dimension matrices for the HQET operators}\label{sec:adms}
In this section we list the anomalous dimension matrices for the $\Delta B = 0$ operators within HQET.
We computed the anomalous dimension matrix with vanishing external momenta, an infrared (IR) regulator $\tilde{\omega} / 2$ for the HQET propagators and a mass $m$ as IR regulator for the light quark and gluon propagators, keeping the gluon gauge parameter $\xi_g$ symbolic.
The relevant integral is given by
\begin{align}
    I \left(m,n\right) &= \int\frac{\mathrm{d}^d k}{\left(2\pi\right)^d} \frac{1}{\left[k^2-m^2\right]^\alpha \left[2 k \cdot v + \tilde{\omega}\right]^\beta} \notag \\
    &= \frac{\mathrm{i} \left(-1\right)^{-\left(\alpha+\beta\right)} 2^{1-\beta-d}}{\pi^{\frac{d-1}{2}}} \frac{\Gamma\left(2\alpha+\beta-d\right)}{\Gamma\left(\alpha\right) \Gamma\left(\alpha+\beta+\frac{1-d}{2}\right)} \left[-\tilde{\omega}\right]^{d-\left(2\alpha+\beta\right)} \notag \\
    & \quad \times {}_{2}{F}_1 \left(\frac{2\alpha+\beta-d}{2}, \frac{2\alpha+\beta+1-d}{2}; \frac{2\alpha+2\beta+1-d}{2}; -\frac{4\tilde{m}^2}{\tilde{\omega}^2} \right) \,,
\end{align}
with $\tilde{m}^2 := m^2 - \frac{\tilde{\omega}^2}{4} - \mathrm{i} \epsilon$ and $\mathrm{Re}\left(d\right) < 2 \alpha + \beta$.
Of course, for the calculation of the one-loop anomalous dimension matrix only the UV pole is needed, which permits the neglect of the mass in the numerator of the fermion propagator, since it does not contribute to the pole.
For the system $\left\{\tilde{O}_{i=1,\ldots,8}^q, \tilde{E}_{i=1,\ldots,6}^q \right\}$ the anomalous dimension matrices are given by
\begin{equation}\label{eq:HQETADMQQ}
    \hat{\tilde{\gamma}}_{\tilde{O}\tilde{O}}^{(0)} = \begin{pmatrix} 
    \frac{3}{N_c} - 3N_c & 0 & 6 & 0 & 0 & 0 & 0 & 0 \\
    0 & \frac{3}{N_c} - 3N_c & 0 & 6 & 0 & 0 & 0 & 0 \\
    \frac{3}{2} - \frac{3}{2N_c^2} & 0 & -\frac{3}{N_c} & 0 & 0 & 0 & 0 & 0 \\
    0 & \frac{3}{2} - \frac{3}{2N_c^2} & 0 & -\frac{3}{N_c} & 0 & 0 & 0 & 0 \\
    0 & 0 & 0 & 0 & \frac{3}{N_c} - 3 N_c & 0 & 0 & 0 \\
    0 & 0 & 0 & 0 & 0 & \frac{3}{N_c} - 3 N_c & -2 & 8 \\
    0 & 0 & 0 & 0 & 0 & 0 & \frac{3}{N_c} - 3 N_c & 0 \\
    0 & 0 & 0 & 0 & \frac{1}{2} \left(-1 + \frac{1}{N_c^2}\right) & 2 - \frac{2}{N_c^2} & \frac{2}{N_c} - N_c & -\frac{5}{N_c} + N_c
    \end{pmatrix} \,,
\end{equation}
and
\begin{equation}\label{eq:HQETADMQE}
    \hat{\tilde{\gamma}}_{\tilde{O}\tilde{E}}^{(0)} = \begin{pmatrix}
    0 & 0 & -\frac{1}{2} & 0 & 0 & 0 \\
    0 & 0 & 0 & -\frac{1}{2} & 0 & 0 \\
    \frac{1}{8} \left(-1 + \frac{1}{N_c^2}\right) & 0 & -\frac{N_c^2 - 2}{4N_c} & 0 & 0 & 0 \\
    0 & \frac{1}{8} \left(-1 + \frac{1}{N_c^2}\right) & 0 & -\frac{N_c^2 - 2}{4N_c} & 0 & 0 \\
    0 & 0 & 0 & 0 & 0 & -\frac{1}{2} \\
    0 & 0 & 0 & 0 & 0 & 0 \\
    0 & 0 & 0 & 0 & \frac{1}{8} \left(-1 + \frac{1}{N_c^2} \right) & -\frac{N_c^2 - 2}{4N_c} \\
    0 & 0 & 0 & 0 & 0 & 0
    \end{pmatrix} \,.
\end{equation}
For the corresponding system of primed operators the anomalous dimension matrices are identical.
The upper-left $4 \times 4$ blocks correspond to the basis of SM HQET operators and have been computed in Refs.~\cite{Neubert:1996we,Ciuchini:2001vx,Kirk:2017juj}; we find agreement with these previous results.
The remaining entries are new.
We note that the slightly more complicated structure of the last row of \cref{eq:HQETADMQQ} is due to the reduction of the tensor Dirac structures by means of \cref{eq:eomtensorHQET}.

\section{Calculation of the mixed quark-gluon condensate contribution}\label{sec:condensateexample}
To calculate the contribution from the mixed condensate $\langle g_s\bar{q}\sigma_{\mu\nu}G^{\mu\nu}q\rangle$, we consider the left bottom diagram in \cref{fig:Cond_FD} and start again with the usual definition of the correlator
\begin{equation}
    \begin{aligned}
        K(\omega_1,\omega_2) = \int &\mathrm{d}^d x_1\,\mathrm{d}^d x_2\;e^{\mathrm{i} p_1 x_1 -\mathrm{i} p_2 x_2} \\
    &\times \langle0|\bar{q}(x_2)\gamma_5h(x_2)\;\bar{h}(0)\Gamma q(0)\otimes\bar{q}(0)\Gamma^\prime h(0)\;\bar{h}(x_1)\gamma_5q(x_1)|0\rangle.
    \end{aligned}
\end{equation}
However in this case, while we still contract the heavy quark fields, we only contract one pair of quark fields to form a light quark propagator with a gluon emission and treat the other two light quarks as soft external states, i.e.
\begin{equation}\label{eq:corrmixcond}
    \begin{aligned}
        K(\omega_1,\omega_2) = \int &\mathrm{d}^d x_1\,\mathrm{d}^d x_2\;e^{\mathrm{i} p_1 x_1-\mathrm{i} p_2 x_2}\\
        &\times \langle0|\wick[sep=5pt]{\c2{\bar{q}}(x_2)\gamma_5\c1{h}(x_2)\;\c1{\bar{h}}(0)\Gamma \c2{q}(0)}\otimes\wick[sep=5pt]{{\bar{q}}(0)\Gamma^\prime \c1{h}(0)\;\c1{\bar{h}}(x_1)\gamma_5{q}(x_1)}|0\rangle.
    \end{aligned}
\end{equation}

When writing the quark propagators, we would like to convert to momentum space as usual, however on the right-hand side of the operator insertion in \cref{eq:corrmixcond}, we have only one propagator and consequently no closed loop.
Introducing the momentum representation on the right-hand side and performing the $\mathrm{d}^d x_1$ integral then forces the heavy-quark propagator to become $1/(2 \omega_2)$ by means of $p_2 \cdot v = \omega_2$.

With the HQET and NLO light quark propagators respectively defined
\begin{align}
    \mathrm{i} S^h_{ij}(k) &= \frac{\mathrm{i}\delta_{ij}(1+\slashed{v})}{2(k\cdot v)}\,, \\
    \mathrm{i} S^{(1)}_{ij}(k) &= -\mathrm{i}g_s\frac{T^b_{ij} G_{\mu\rho}^b(0)}{4(k^2-m^2)^2}\left((\slashed{k}+m)\sigma_{\mu\rho}+\sigma_{\mu\rho}(\slashed{k}+m)\right)\,, \label{eq:fermionpropexpandedfirstorder}
\end{align}
we write the full correlator for the $\tilde{T}_1$ operator as 
\begin{equation}
    \begin{aligned}
        K_{\tilde{T}_1}(\omega_1,\omega_2) = \textbf{Colour} &\times \int \frac{\mathrm{d}^d k_2}{(2\pi)^d}\frac{\mathrm{Tr}\left[(\slashed{k}_2 \sigma_{\mu\nu}+\sigma_{\mu\nu}\slashed{k}_2)\gamma_5 (1+\slashed{v})\gamma_\sigma(1-\gamma_5)\right]}{4 \left(k_2^2\right)^2 \cdot 2(k_2\cdot v+\omega_2)} \\
                                            &\times \left\{\frac{\left[\gamma^\sigma(1-\gamma_5) (1+\slashed{v})\gamma_5\right]_{\epsilon\lambda}}{2(k_1\cdot v)}\right\}_{k_1=p_1} \langle0|\bar{q}_{l\epsilon}(0)q_{n\lambda}(x_1)|0\rangle.
    \end{aligned}
\end{equation}
Taking the quark field expansion at leading order, i.e.\ $q(x_1)=q(0)$, the sought-after condensate can be constructed after some colour algebra:
\begin{equation}
    \begin{aligned}
        \textbf{Colour}\cdot\langle\dots\rangle &= g_s\,T^b_{ik}T^a_{jk}T^a_{lm}\delta_{ij}\delta_{mn}\,G^b_{\mu\nu}\,\langle0|\bar{q}_{l\epsilon}(0)q_{n\lambda}(x_1)|0\rangle \\
                                                     &= g_s\frac{\delta^{ab}}{2}T^a_{ln}\,G^b_{\mu\nu}\,\langle0|\bar{q}_{l\epsilon}(0)q_{n\lambda}(0)|0\rangle \\
                                                     &= \frac{\delta_{ll}}{24d(d-1)}\langle g_s\bar{q}\sigma_{\kappa\tau}G^{\kappa\tau}q\rangle\left(\sigma_{\mu\nu}\right)_{\lambda\epsilon},
    \end{aligned}
\end{equation}
where the last step uses the relation~\cite{Pascual:1984zb}
\begin{equation}
    \langle0|\bar{q}_{i\alpha}(0)\,T^a_{ij}G^{a\mu\nu}\,q_{j\beta}(0)|0\rangle = \frac{\delta_{ii}}{12d(d-1)}\langle \bar{q}\sigma_{\kappa\tau}G^{\kappa\tau}q\rangle \left(\sigma_{\mu\nu}\right)_{\beta\alpha}.
\end{equation}
Inserting this into the correlator and performing the Dirac traces, we find
\begin{equation}
    \begin{aligned}
        K_{\tilde{T}_1}(\omega_1,\omega_2) = \frac{N_c}{96d(d-1)} &\times \frac{1}{2(k_1\cdot v)}\bigg|_{k_1=p_1} \, \langle g_s\bar{q}\sigma_{\kappa\tau}G^{\kappa\tau}q\rangle \\
                                           &\times \int \frac{\mathrm{d}^d k_2}{(2\pi)^d}\frac{-192(v\cdot k_2)}{\left(k_2^2\right)^2\, 2(k_2\cdot v+\omega_2)}.
    \end{aligned}
\end{equation}
This expression can be further reduced, leading to 
\begin{align}
    K_{\tilde{T}_1}(\omega_1,\omega_2) &= \frac{N_c}{2\omega_1} \frac{\langle g_s\bar{q}\sigma Gq\rangle(d-3)}{\omega_2 d(d-1)}I(\omega_2),
\end{align}
where the one-loop integral is defined~\cite{Grozin:2003ak}
\begin{align}
    I(\omega) &= \int\frac{\mathrm{d}^d k}{(2\pi)^d}\frac{1}{2(k\cdot v+\omega)\,k^2} = \frac{\mathrm{i}}{(4\pi)^{d/2}}\frac{\Gamma(1-\epsilon)\Gamma(-1+2\epsilon)}{\Gamma(1)}(-2\omega)^{1-2\epsilon}.
\end{align}
Note that while this correlator has a discontinuity for positive, real $\omega_2$, it is not discontinuous in $\omega_1$ where it instead has a simple pole at $\omega_1=0$.
In order to express the dispersion relation as an integral over the positive real line such that it can be used in the sum rule, we write
\begin{equation}
    \oint d\nu\,\frac{A}{\nu(\nu-\omega)} = \int_0^\infty d\nu\,\frac{A}{\nu-\omega}\delta(\omega),
\end{equation}
where $A$ is some constant.
The spectral density contribution for the mixed quark-gluon condensate therefore reads
\begin{equation}
    \Delta\rho^{\text{cond}}_{\tilde{T}_1} = \frac{N_c\langle g_s\bar{q}\sigma Gq\rangle}{192\pi^2}\left[\delta(\nu_1)+\delta(\nu_2)\right],
\end{equation}
where we have further included the $\omega_1\leftrightarrow\omega_2$ symmetric diagram on the bottom right in \cref{fig:Cond_FD}.

\bibliographystyle{JHEP-jmf-arxiv}
\bibliography{lit.bib}

\end{document}